\newcommand{\name}{self-evolving computing system}
\newcommand{\Name}{Self-evolving computing system}
\newcommand{\NAME}{Self-Evolving Computing System}
\newcommand{\names}{self-evolving computing}
\newcommand{\Names}{Self-evolving computing}
\documentclass[letterpaper,11pt,twoside,onecolumn,final]{JIDPS}       
\usepackage{titlesec}								
\titlelabel{\thetitle.\quad}							
\usepackage[natbibapa]{apacite}						
\usepackage[margin=1in]{geometry}   				
\usepackage{mathptmx}				 				
\usepackage{amsmath}					 			
\usepackage{multirow} 					 			
\usepackage{epstopdf}					 			
\usepackage[dvips]{graphicx}             	 			
\usepackage{subfigure}                     					
\graphicspath{{figures/}}                    				
\usepackage{algorithmicx, algpseudocode}			
\usepackage{enumerate}
\usepackage{indentfirst}                     				
\usepackage{setspace}                       				
\usepackage{algorithm}						   		
\usepackage{lineno,xcolor}							
\usepackage{fancyhdr}                       				
\usepackage[font={footnotesize,bf,sf}, labelfont={sf,bf}, justification=centering, labelsep=period, margin=5mm]{caption}	 
\usepackage{times}								
\usepackage{everypage}                      			
\usepackage{ifthen}                             			
\usepackage[bookmarksopen,bookmarksnumbered,colorlinks,urlcolor=blue,linkcolor=blue,anchorcolor=blue,citecolor=blue]{hyperref} 	
\usepackage{verbatim}

\newenvironment{AuthorThanks}
  {\par\edef\savedfootnotenumber{\number\value{footnote}}
   
   \setcounter{footnote}{0}}
  {\par\setcounter{footnote}{\savedfootnotenumber}}

\AddEverypageHook{
  \ifthenelse{\value{page}=1}
    {\setlength{\headheight}{65pt} \setlength{\headsep}{-40pt} } 
    {\setlength{\headheight}{13pt} \setlength{\headsep}{25pt} }} 

\begin{document}

\vspace*{24mm}

\begin{spacing}{2}

\noindent{\LARGE \bfseries
The Vision of \NAME{}s
}

\end{spacing}
\vspace*{4mm}

\begin{AuthorThanks} 
\noindent{Danny Weyns, Katholieke Universiteit Leuven, Belgium and Linnaeus University, Sweden
	\footnote{Corresponding author. Email: 
		\href{mailto: danny.weyns@kuleuven.be}{danny.weyns@kuleuven.be}. Tel: (+32)474-208251.
	}
}\\
\noindent{Thomas B\"ack, Leiden University, The Netherlands and NORCE Norwegian Research Centre, Norway}\\
\noindent{Ren\`e Vidal, Johns Hopkins University, USA and NORCE Norwegian Research Centre, Norway}\\
\noindent{Xin Yao, University of Birmingham, UK and Southern University of Science and Technology, China}\\
\noindent{Ahmed Nabil Belbachir, NORCE Norwegian Research Centre, Norway}
\end {AuthorThanks}

\vspace*{2mm}

\vspace*{10mm}

\noindent\textbf{Abstract} {
Computing systems are omnipresent; their sustainability has become  crucial for our society. A key aspect of this sustainability is the ability of computing systems to cope with  the continuous change they face, ranging from dynamic operating conditions, to changing goals, and technological progress. While we are able to engineer smart computing systems that autonomously deal with various types of changes, handling unanticipated changes requires system evolution, which remains in essence a human-centered process. This will eventually become unmanageable.~To break through the status quo, we  put forward an arguable opinion for the vision of \textit{\name{}s} that are equipped with an evolutionary engine enabling them to evolve autonomously. Specifically, when a \name{} detects conditions outside its operational domain, such as an anomaly or a new goal, it activates an evolutionary engine that runs online experiments to determine how the system needs to evolve to deal with the changes, thereby evolving its architecture.~During this process the engine can integrate new computing elements that are provided by computing warehouses.~These computing elements provide specifications and procedures enabling their automatic integration. We motivate the need for \name{}s in light of the state of the art, outline a conceptual architecture of \name{}s, and illustrate the architecture for a future 
smart city mobility system that needs to evolve continuously with changing conditions.~To conclude, we highlight key research challenges to realize the vision of \name{}s. 
}

\vspace*{6mm}



\section{Introduction}
Our society is going through a digitization process that penetrates virtually every aspect of our life, from health and industries, to transportation, public services, and entertainment. Consequently, we increasingly depend on the sustainability of computing systems.~Yet, achieving this sustainability is  challenging~\citep{DBLP:conf/sfm/2007,LEHMAN200333,EU} and spans manifold areas, from quality of service and software evolution to energy-awareness and software engineering processes.~One key aspect to achieve sustainability of computing systems is managing the complexity that arises from the ever changing conditions these systems face. Such changes may or may not be anticipated when the system was built and include dynamics in the environment, new emerging goals,\footnote{We use goals and requirements interchangeably in this paper.} and the introduction of new technologies. We take this angle of change to sustainability of computing systems. 

Currently we can build smart computing systems that can deal with many tasks autonomously, adapt themselves or learn over time to deal with changes. Other tasks can be managed by system operators, for instance, perform predictive maintenance. However, current computing systems can only handle changes that were anticipated, that is, changes that occur within the  operational domain for which the system has been built. Current smart computing systems cannot handle unanticipated changes, such as anomalies outside their operational domain, and the emerge of new goals or new technologies. Such changes require  evolution of the computing system.~Although significant progress has been made on automating the deployment and integration of new elements, software evolution remains in essence a human-driven activity. 

With the ever increasing complexity of computing systems and the continuous changes these systems are subjected to, human-driven approaches will eventually become unmanageable~\citep{10.1145-336512.336534,10.1109-FOSE.2007.20,Reussner2019,1882362.1882367,Andersson2013}. The capacity to handle large amounts of data and the availability of efficient decision algorithms opens perspectives to major breakthroughs towards fully autonomous systems that operate in continuous changing environments~\citep{DNV,WeynsBCCFGNPRRS21,Weyns2022}. However, we currently lack fundamental knowledge to turn these long-standing challenges into reality. 

When comparing the capabilities of present-day computing systems with those of biological systems a few striking conclusions can be drawn. In contrast to computing systems, biological systems have a remarkable ability to deal with changes. For instance, insects have exceptionally fast reactions and can avoid dangerous situations or locate hidden food sources by swiftly \textit{adapting} to their environment~\citep{Camazine:2003}. They have also \textit{evolved} dramatically, from one generation to the next, to accommodate changes over time in their habitat and the climate conditions. 

\begin{figure*}[!thb]
    \centering
    \includegraphics[width=0.86\linewidth]{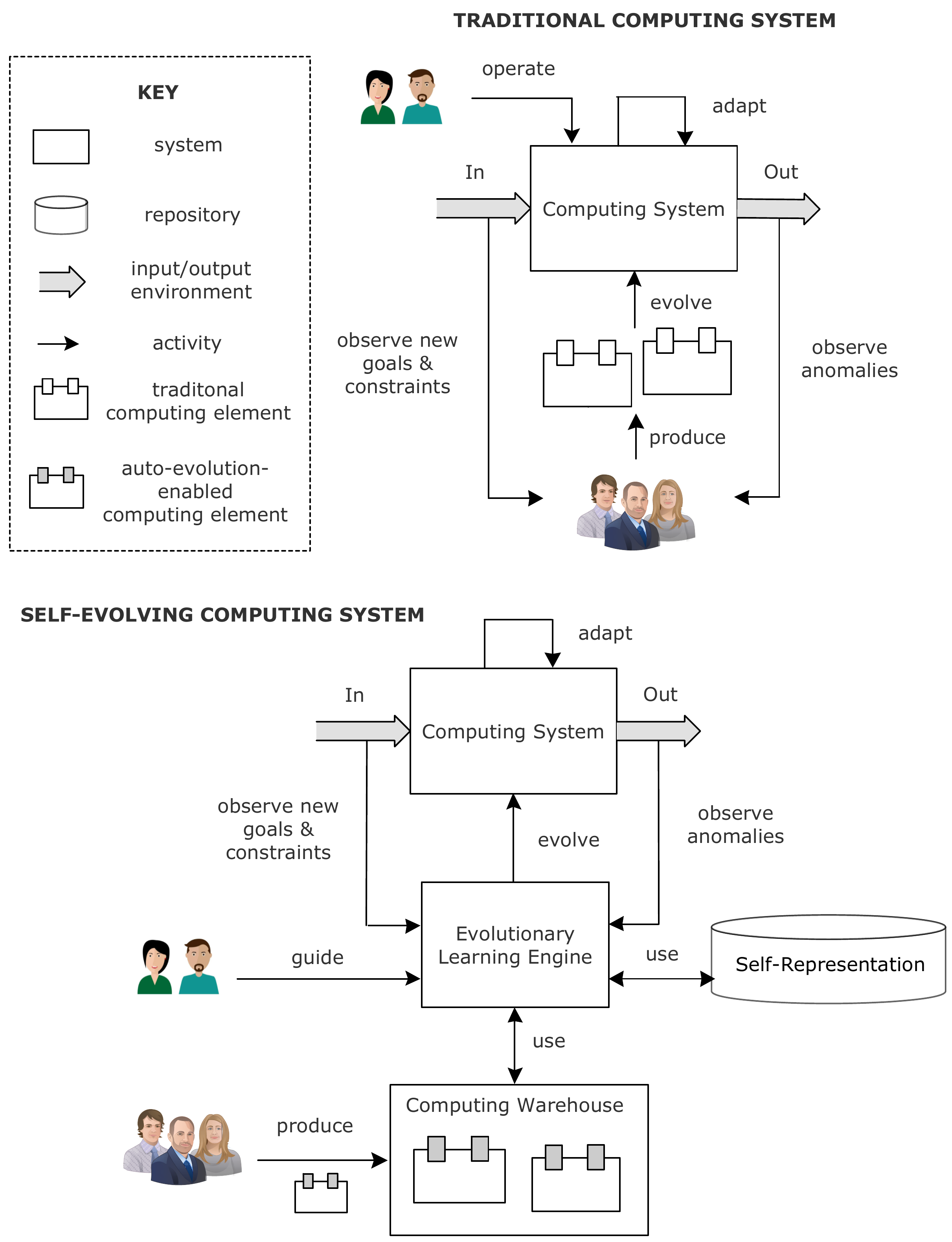}\vspace{10pt}
    \caption{From traditional computing systems to \name{}s. 
    }
    \label{fig:overview}
\end{figure*}

Inspired by the principles of biological systems, this paper puts forward an arguable opinion for the vision of \textit{\name{}}s, i.e., computing systems that evolve themselves autonomously. 
Figure~\ref{fig:overview} illustrates how \name{}s differ from traditional computing systems. 
A \textit{traditional computing system} takes inputs from the environment and produces outputs in the environment, realizing the users' goals~\citep{1997Jackson}.
To deal with changing conditions, such a system can be equipped with smart techniques, either internally (e.g., a learning algorithm) or externally via a feedback loop, enabling the system to \textit{self-adapt} its configuration autonomously to deal with changes~\citep{Rainbow,weyns2020book}. 
A traditional computing system is designed to work in an operational domain, i.e., well-defined conditions of the environment in which the system should achieve its goals. Humans may be involved to \textit{operate} the system, for instance to start/stop the execution of batches of tasks or to perform predictive maintenance. Extending the operational domain, for instance to deal with \textit{new goals or new constraints}, or to mitigate \textit{anomalies}, requires the system to undergo an \textit{evolution} step that typically relies on humans that produce new computing elements that are then \textit{deployed and integrated} into the system, a process that is increasingly automated~\citep{RODRIGUEZ2017263}. 

In contrast, a \textit{\name{}} maintains a \textit{self-representation} that includes runtime models of the computing system and its goals (self-awareness), 
and the environment in which the system operates (context-awareness). 
An \textit{evolutionary learning engine} uses the self-representation to autonomously evolve the architecture of the computing system, in response to \textit{unanticipated changes} that occur throughout the system's lifetime, i.e., new goals or new constraints that appear, or anomalies identified during operation. 
 To that end, the evolution engine runs experiments in a sandbox evolving the system model until it satisfies the new conditions. During this process, the engine can integrate new computing elements from \textit{computing warehouses} as needed. These \textit{auto-evolution-enabled} computing elements provide specifications and procedures that enable an evolutionary engine to incorporate these elements autonomously. As shown in Figure~\ref{fig:overview}, a \name{} takes the human out of the loop of the evolution process. Humans only produce new auto-evolution-enabled computing elements that are readily available for \name{}s via computing warehouses. Yet, humans may be involved to provide \textit{guidance} to the system, for instance to set constraints on the behavior of the system or express preference of one configuration over another during evolution.  \Name{}s focus on the evolution aspects of computing systems within the newly proposed paradigm of ``lifelong computing''~\citep{lifelong}. \Name{}s also resemble similarities with the idea of  ``self-growing software'' proposed by~\cite{Tamai2019} as the next paradigm shift in software engineering.  

The remainder of this paper starts with a discussion of a selection of key approaches to deal with change and points out why a novel foundation is required  (Section~\ref{sec:soa}). 
Then we introduce an illustrative example (Section~\ref{sec:example}). We outline a conceptual architecture for \name{}s 
(Section~\ref{sec:architecture}) and illustrate the architecture for the example. To conclude, we highlight key research challenges for realizing the vision of \name{}s and we suggest starting points to tackle them (Section~\ref{sec:conclusions}).

\section{State of the Art}\label{sec:soa}

Already in the early 2000s, IBM pointed to the manageability problems caused by the growing complexity of computing systems~\citep{2003IBM}. In response, they launched the autonomic computing initiative that was centered on enabling computing systems to manage themselves based on high-level goals, similar to the autonomic nervous system of the human body. Autonomic computing primarily focuses on automating tasks of running computing systems that are traditionally done by operators. Hence, the target of autonomic computing is the operational domain of computing systems. Self-evolution on the other hand targets the \textit{autonomous evolution} of computing systems, hence the target is a change of the operational domain. Self-evolution aims to enable computing systems dealing with unanticipated change by evolving autonomously. 

In this section, we summarize the state of the art in two key fields that tackle the problem of managing change of computing systems from two complementary points of view: smart systems and software evolution. Based on this analysis, we motivate the need for self-evolving systems.  

\subsection{Smart Systems}

\cite{8329014} surveyed smart computing systems, with an emphasis on cyber-physical systems. The authors distinguish four levels of smartness mapping to increasingly challenging types of changes to be tackled by the systems, ranging from no changes to unknown changes. Smartness then refers to the capability level of computing systems to handle these types of changes through reasoning, learning, adapting, and evolving. \cite{Weyns2022} extended the notion of smart to ``smarter''  referring to both computing systems and their engineering processes that continuously adapt and evolve through a perpetual process that continuously improves their capabilities and utility to deal with the uncertainties and new data they face throughout their lifetime. 
\cite{3089649.3089656} emphasized that smartness of computing systems enable them to deal with dynamics and uncertainty in the environment, and external threats. The authors highlight that smartness of computing systems is primarily implemented through the software leveraging principles from self-adaptation.~\cite{Musil2017} presented a set of architectural patterns to realize self-adaptation across the software stack of cyber-physical systems. 

A classic field of study on smartness is autonomous systems (or intelligent autonomous systems) \citep{Tzafestas2012,Paulovich2018}.~Autonomous systems mimic human (or animal) intelligence, in order to operate independently of direct human supervision. An important sub-field of autonomous systems is multi-agent systems~\citep{Wooldrige2009} that studies the operation and coordination of autonomous agents that aim at solving problems that go beyond the capabilities of single agents. Different authors have presented patterns that document problem-solution pairs for engineering multi-agent systems~\citep{SchelfthoutKurt2002Aip,Dastani16,Marks18}. \cite{JuziukJoanna2014DPfM}~presented a systematic literature overview classifying patterns based on focus, granularity, level of abstraction, and source of inspiration. The field of human-robot teams~\citep{MUSIC201642} studies collaboration of humans and robots exploiting their complementary skill sets. 
Another promising key field enabling the realization of smartness is digital twins~\citep{8477101}. Digital twins are characterized by the seamless integration between the cyber and physical spaces. Digital twins have been successfully applied in product design, production, prognostics and health management, among other fields. 
Gentelligent systems~\citep{Denkena:2017} integrate sensing components throughout the production supply chain to improve efficiency, flexibility, and product quality. 
Recently, the interest in autonomous systems has been expanding significantly with high-profile applications, such as smart robotics (Industry 4.0 driven by the Internet of Things) and smart transportation. 
For instance, \cite{6857843} stressed the need to equip Industry 4.0 systems with smart actuators, sensors, and telecommunication technologies, providing these systems access to the higher-level processes and services. \cite{978-3-030-00761-4} presented MARTAS that automates the management of Internet-of-Things leveraging statistical model checking at runtime to ensure the system goals under uncertainty. 
\cite{7433937} referred to smartness of the electricity grid as the integration of information and communication technology with other advanced technologies that enable electric energy generation, transmission, distribution, and usage to be more efficient, effective, economical, and environmentally sustainable. 
\cite{8010538} investigated smart transportation systems using a modeling and simulation environment. Smartness in this context relates to the ability of a system to deal with attacker-defender behavior, including vulnerability analysis to traffic signal tampering, resilient sensor selection for forecasting traffic flow, and resilient traffic signal control in the presence of denial-of-service attacks. 

Another classic field of smart systems is self-adaptation.~Simultaneous with industrial initiatives, such as autonomic computing~\citep{Kephart} mentioned above, researchers studied the abilities of computing systems to handle change autonomously~\citep{Oriezy1999,Rainbow}. Self-adaptation is based on the principles of feedback computing~\citep{Oriezy1999,2007Kramer,1516538,weyns2020book}. 
Over the past two decades, extensive efforts have been put in devising fundamental principles of self-adaptation as well as techniques and methods to engineer self-adaptive systems~\citep{DBLP:books/sp/19/Weyns19}. Whereas the initial focus was on automating operator tasks based on high-level goals~\citep{Kephart,Rainbow}, later research shifted towards taming uncertainties that computing systems face during operation and that are difficult to anticipate before deployment~\citep{Cheng2009,2786805.2786853,8008800}. This view introduces a perspective that blends  system  engineering and system operation~\citep{1882362.1882367,10.1007/978-3-319-74183-3_2,10.1145-3204459,10.1145-3190507}. Central to any self-adaptive systems are runtime models~\citep{Blair2009} that provide the system with self-awareness (self-representation and representation of goals) and context-awareness (representation of the environment)~\citep{9089006,10.1145-3347269,FORMS}. These models are updated at runtime tracking uncertainties~\citep{Esfahani2013,MAHDAVIHEZAVEHI201745,10.1145/2797433.2797497,9196226} and then used to analyze the situation and decide when and how to adapt the system to maintain its goals, or gracefully degrade if needed.  

\subsection{Software Evolution}

Evolution is a natural part of the life cycle of software systems that traditionally occurs in incremental development in response to changes in the environment, purpose, or use of the software system~\citep{Reussner2019}. 
\cite{10.5555-1090744.1090746} presented a taxonomy for software evolution with four dimensions of system change: temporal properties (i.e., when do changes happen), objects of change (i.e., where in the system do we make changes), system properties (i.e., what is changed), and change support (i.e., how is the system changed). Earlier, \cite{10.5555-371697.371701} identified two other core dimensions: motivations (i.e., why are the changes done) and roles (i.e., who is doing system changes). The ISO/IEC standard for software maintenance\footnote{International Organization for Standardization. ISO/IEC 14764. 2014. URL: \url{www.iso.org/standard/39064.html}} distinguish four types of software changes: corrective (bug fixing dealing with errors), adaptive (environment and requirement changes), perfective (optimizing or refactoring the system), and preventive modifications (preventing problems).

During the past decades, the traditional view of software that evolves through periodic releases has been replaced by continuous evolution of software~\citep{RODRIGUEZ2017263}. Software organizations today develop, release, and learn from software in rapid parallel cycles (typically from hours to a few weeks). This approach is commonly referred as continuous deployment (CD)~\citep{978-3-319-08738}. CD is based on the principles of agile development~\citep{DINGSOYR20121213} and DevOps~\citep{MISHRA2020100308} that aim at increasing the deployment speed and quality of systems. CD leverages on continuous integration (CI)~\citep{6802994} that automates tasks such as compiling code, running tests, and building deployment packages. Among the benefits of CI/CD are rapid innovation, shorter time-to-market, increased customer satisfaction, continuous feedback, and improved developer productivity. Yet, an important concern of current practice in software maintenance is (intentional or unintentional) technical debt, i.e., longer-term negative effects on systems that result from sub-optimal decisions~\citep{LI2015193}, in particular in the context of agile development. Furthermore, researchers have argued that the current level of automation needs to be enhanced ~\citep{RODRIGUEZ2017263}, and last but not least, to develop sustainable computing systems, we need  sustainable software development processes~\citep{GREENSOFT,Dick2010,Andersson2013,Weyns2022,10.1145/3337773,weyns2019activforms}.

With the increasing exposure of computing systems to change, the volumes of data they need to process, and the seamless integration of humans in the loop~\citep{6008519,Selic20,ZENG20201028,7158500}, computing systems face uncertainties that are difficult or even impossible to predict before deployment. Hence, engineers may not be able to obtain sufficient knowledge to make all design decisions before the system is deployed. This calls for postponing design decisions until after deployment when the required knowledge becomes available. The design decisions are then enacted through continuous adaptation and evolution~\citep{1882362.1882367,weyns2020book}. To that end, a number of important building blocks have been studied. We highlight two: anomaly detection and lifelong learning.  

Anomaly detection (or outlier or novelty detection) aims at identifying data instances that significantly deviate from the majority of data instances in a data set~\citep{1969Grubbs}. Anomaly detection  has been used in a variety of domains, e.g., intrusion detection, fault prevention, defect detection, and unexpected flow  detection. A plethora of methods have been developed~\citep{3381028,6515601}, including proximity-based approaches that rely on relations between nearby data points, projection techniques that convert data into a space with reduced dimensionality to improve outlier detection, outlier detection for multi-dimensional data such as recursive binning and re-projection, windowing for online time series that incrementally builds and updates models with new data, learning model spaces for fault diagnosis, and deep learning anomaly detection, such as deep neural network auto-encoders. Yet, dealing with highly complex data remains an open problem. Anomaly detection mechanisms enable a computing system to autonomously identify behavior at the boundaries or outside its operational domain, providing a basis building block for the realization of self-evolving systems.  

Lifelong learning (or continual learning) refers to the ability of a system to continually accommodate new knowledge to learn new tasks that were not predefined~\citep{978-3-642-79629-6-7}.~Different approaches for lifelong learning have been developed relying on supervised, unsupervised, and reinforcement learning~\citep{LML}, and  recently lifelong learning based on neural networks has gaining increasing interest~\citep{PARISI201954}. 
A key challenge for lifelong learning is dealing with catastrophic forgetting that refers to the loss of previous learning while learning new information; this may lead to failures for systems operating in real-world environments~\citep{HASSELMO2017407}. Different approaches have been proposed to deal with this problem, such as dynamic allocating new neurons or network layers to accommodate novel knowledge, and using complementary learning networks with experience replay, yet more research is needed apply these techniques to real-world systems~\citep{PARISI201954}. Lifelong learning techniques provide another basic block for the realization of \name{}s. 
 
 \subsection{Why \NAME{}s?}

When we look at the current landscape of research, we can observe two principle lines of work. The first line studies the application of smart techniques enabling systems to deal with changes autonomously during operation. The second line studies the evolution of computing systems with an emphasizes on tools for automating the deployment and integration of computing elements. We advocate that a key underlying problem with these existing approaches is the lack of an integrated perspective on handling change---anticipated and unanticipated---in an autonomous manner. Compared to traditional (or conventional) systems, smart systems are equipped with capabilities to handle a variety of changes autonomously. Yet, the target domain of smart systems is in essence their operational domain, that is, their capabilities are confined to what they have been built for. The aim of software evolution lays essentially in revising or extending the operational domain. While several steps in the process of software evolution have been automated in the past decades, the actual evolution of the software remains in essence a human-driven activity. Autonomous and self-adaptive systems have expanded the operational domain of computing systems substantially, enabling them to deal with changes during operation to enhance their efficiency and being most robust, yet the scope remains bounded to anticipated changes. Anomaly detection mechanisms allow identifying deviations from expected behaviors, and lifelong learning enables learning-based systems dealing with new tasks during operation. Yet, besides their current limitations for real-world problems, these techniques offer only basic blocks to realize a true integration of continuous adaptation and evolution aiming at mitigating the effects of uncertainty that spans both anticipated and unanticipated change. To tackle the challenges of continuous change, anticipated and unanticipated, a new integrated perspective for the engineering and operation of future computing systems is needed. \Name{}s aim to offer such a perspective.

\section{Future Smart City Mobility Scenario}\label{sec:example}  

We illustrate the need for \name{}s with an example of a future smart city mobility scenario. A research study called ``New autoMobility''~\citep{acatech2015} highlighted how automated and networked vehicles and trains can be usefully integrated into a user-friendly, efficient and sustainable mobility system in the future. Such a system would consist of mobility hubs, car sharing and self-parking vehicles, and autonomous trains. Flexible, time-and-space-dependent mobility pricing will ensure more evenly distributed usage of mobility resources and prevent traffic gridlock. Vehicles will be able to warn each other (directly or indirectly) in dangerous situations creating a cooperative mixed traffic. Such intelligent, networked transport protects the environment and the climate and improves quality of life.

Establishing automated mobility requires a phased introduction and continuous evolution of a mobility platform to align with a variety of changes.~This poses difficult often conflicting challenges, spanning business, technical, social, and legal aspects. For example, the introduction of automated traffic will happen only gradually, so initially automated and conventionally controlled vehicles will co-exist. 
Depending on local conditions, there may be a need to manage the level of pollution in areas with more intensive traffic of conventionally controlled vehicles. This may require the need for tracking the levels of pollution in these areas and take measures when needed. Such measures may range from temporally redirecting conventional vehicles in certain areas using smart traffic boards, up to increasing prices for polluting vehicles for instance to part in sensitive areas. However, with changing usage profiles, transitions to automated mobility, and novel technological advances, these provisions will need to evolve. 

Central to the functionality and safety of mobility will be the collection and processing of data and information from various sources. Managing this data requires a suitable framework that creates connectivity between vehicles, the infrastructure, and traffic management systems, ensuring safety while respecting the personal interests and privacy concerns of the users at any time. Tackling these challenges and balancing the trade-offs between the various needs will require an integrated computing system that is capable to operate, adapt, and evolve autonomously throughout its lifetime in a continuously changing environment. We illustrate how a \name{} could offer such a unique solution. 

\section{Conceptual Architecture for \NAME{}s}\label{sec:architecture}

In this section, we present a conceptual architecture for \name{}s. 
To deal with the continuous changes a \name{} faces throughout its lifetime, we outline five complementary requirements for a \name{}. These requirements naturally target the ability of self-evolving systems to deal with anticipated change (1), to discover unanticipated changes and evolve autonomously (2-4), and to integrate humans in the loop when needed (5). 

\begin{enumerate}
    \item A \name{} should be able to handle vast amounts of data and realize its goals under changing but anticipated conditions; 
    \item A \name{} should be able to discover and integrate new computing elements autonomously; 
    \item A \name{} should be able to autonomously detect unanticipated conditions, i.e., learn conditions outside its operational domain, including anomalies, new goals and constraints; 
    \item A \name{} should be self-aware and context-aware enabling it to autonomously evolving its architecture to realise its goals; 
    \item Depending on the domain at hand, some activities of a \name{} may be supported by humans. 
    
\end{enumerate}

Requirement (1) is a basic requirement for systems that need to achieve their goals while dealing with huge amounts of data and operating under uncertainty. Requirements (2) to (4) are key for enabling systems to evolve autonomously when encountering unanticipated changes. As for requirement (5), support for human guidance is particularly important: (i) in  domains with critical goals where humans will have the ultimate control over the system by setting boundaries on the system behavior, ensuring the trustworthiness of the system, (ii) for systems that require human interaction to set high level goals or express preferences among possible options generated by the system (in contrast to performing standard operating activities).

\begin{figure}[!thb]
    \centering
    \includegraphics[width=0.6\linewidth]{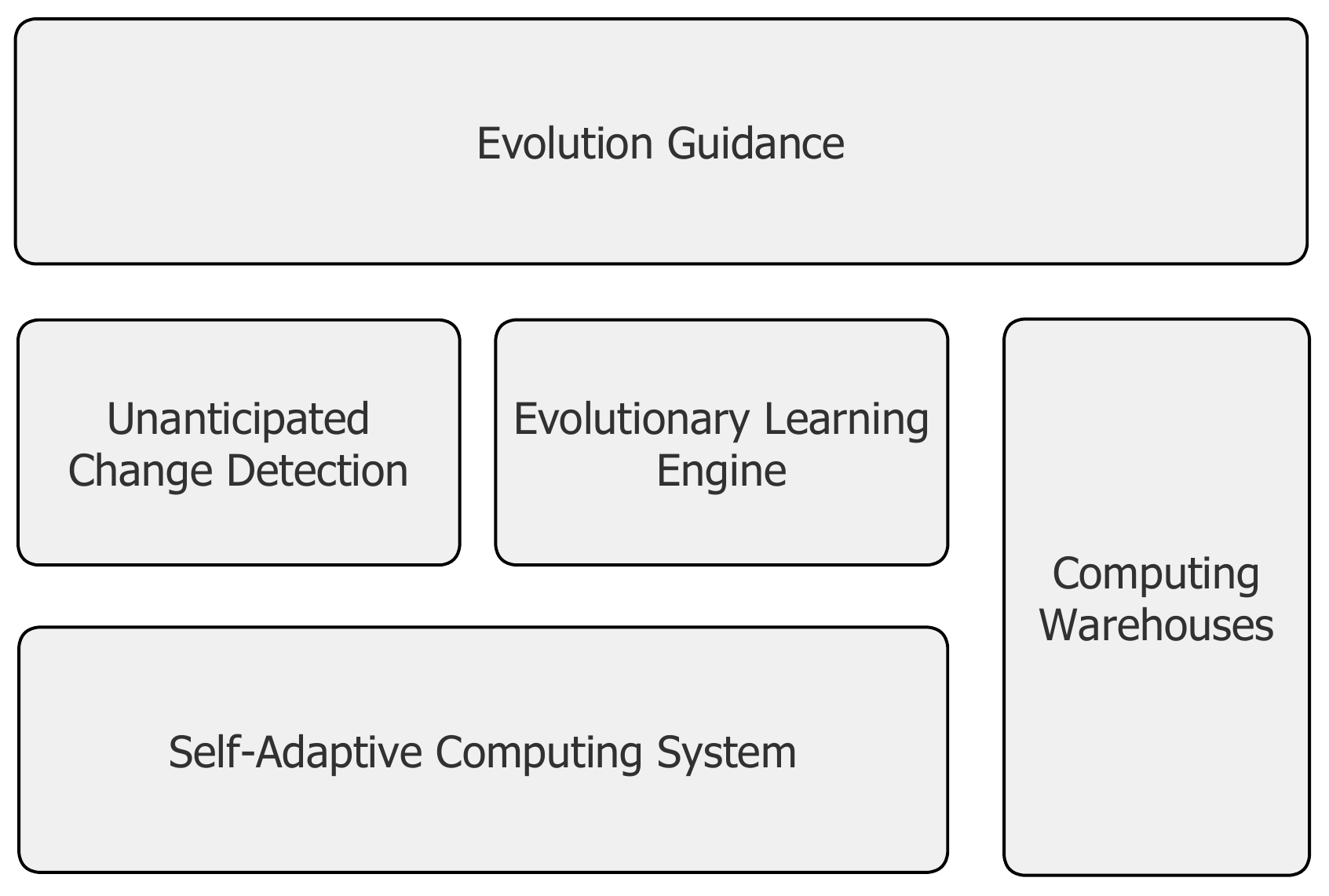}\vspace{8pt}
    \caption{Conceptual architecture for \name{}s with the different building blocks.}
    \label{fig:blueprint}
\end{figure}

To achieve these requirements, we propose a conceptual architecture for \name{}s as shown in Figure~\ref{fig:blueprint}. We explain the different building blocks and illustrate each of them with examples of the future mobility scenario. Starting points to realize the building blocks are explained in Section~\ref{sec:conclusions}.

\vspace{5pt}\noindent\textbf{Self-Adaptive Computing System}. 
As a basis, a \name{} comprises an \textit{self-adaptive computing system} that integrates regular computing elements and learning algorithms, enabling it to handle a vast amount of data and realize the goals of its users. Furthermore, the self-adaptive computing system is equipped with smart techniques enabling it to deal with changes within its operational domain, i.e., changing operation conditions and uncertainties that can be managed by adapting the running architectural configuration of the self-adaptive computing system, without the need for updates or the integration of new computing elements or learning algorithms. As such, a self-adaptive computing system realizes requirement (1). To account for unanticipated changes that requires evolution (see evolutionary learning engine below), the self-adaptive computing system should support automatic updates of its running architecture.

\begin{figure*}[!thb]
    \centering
    \includegraphics[width=\linewidth]{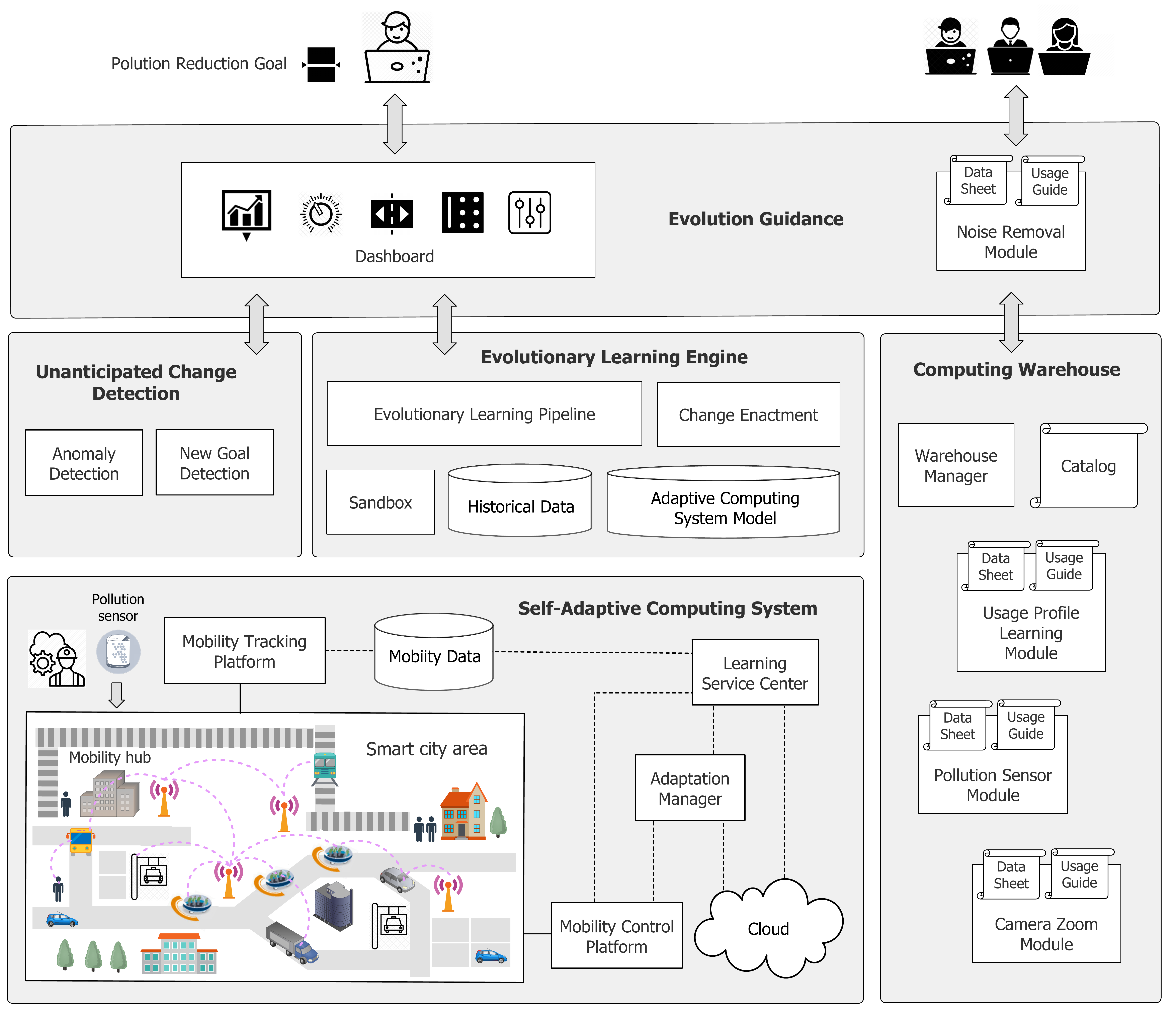}
    \caption{Illustration of the conceptual architecture for a smart city mobility scenario}\vspace{10pt}
    \label{fig:blueprint-detail}
\end{figure*}

Figure~\ref{fig:blueprint-detail} illustrates a self-evolving computing system for the smart city mobility scenario.  
We focus here on the self-adaptive computing system (lower box left) that comprises the smart city area with a mobility hub that connects different modes of public transport, conventional cars and smart vehicles, pedestrians, and a variety of sensors (cameras, smart boards, parking sensors, etc.) that measure the density of traffic, occupation of automated trains, usage of parking lots, movements of pedestrians, etc. The data is collected by a \textit{mobility tracking platform} and stored and updated in a \textit{mobility data repository}. The data is continuously processed by a \textit{learning service center} that learns and predicts relevant system parameters, such as mobility distribution, traffic safety, etc. These parameters together with other data obtained from the Cloud (e.g., weather forecasts) are then used by the \textit{adaptation manager} that continuously optimizes the different objectives of the mobility system and their trade-offs, using the \textit{mobility control platform}. For instance, when a camera detects an increase of passengers of smart vehicles for a particular trajectory, the frequency of these transports may be increased dynamically and the ticket price may be adjusted temporally. 

\vspace{5pt}\noindent\textbf{Computing Warehouses}. 
\Name{}s are supported by \textit{computing warehouses} that offer new computing elements, realizing requirement (2). Computing warehouses leverage the principles of off-the-shelf components and services, open source software, and open data. Computing warehouses can be operated directly by producers of new auto-evolution-enabled computing elements or indirectly via a broker. We refer to the elements provided by computing warehouses as \textit{auto-evolution-enabled computing elements}; examples are a module that offers improved or new functionality, a connector to connect with and use a new external service, a template of new learning algorithm, a repository of data, etc. It is important that \name{}s can incorporate auto-evolution-enabled computing elements autonomously during operation. To that end, each auto-evolution-enabled computing element is equipped with a \textit{data sheet} that specifies its functions, properties, usage requirements, etc., and a \textit{usage guide} that specifies the procedures that need to be followed for using the element. These specifications require both a well-defined syntax and an ontology that defines the semantics of the properties and usage of the elements. Depending on the requirements, new auto-evolution-enabled computing elements may require certification before making them available in a warehouse.  All interactions with the computing warehouse happen via a \textit{warehouse manager}. Clients can search the available elements via a \textit{catalog} that lists the elements with their data sheets and usage guides; using a computing element may be subject to a contract. 

Figure~\ref{fig:blueprint-detail} shows a few examples of new auto-evolution-enabled computing elements for the smart mobile city scenario (box right). The \textit{camera zoom module} provides the software that is required to activate and use zoom lenses on cameras. The \textit{usage profile learning module} offers new learning models of users of a smart city mobility system, possibly derived from studies. The \textit{pollution sensor module} offers the software to start using sensors that measure particular pollution parameters of the environment in the city. 

\vspace{5pt}\noindent\textbf{Unanticipated Change Detection}. 
A key feature of \name{}s is their ability to detect unanticipated changes, i.e., changes that cannot be handled by the build-in learning and adaptation mechanisms of the self-adaptive computing system, realising requirement (3). Such unanticipated changes can be triggered either by an \textit{anomaly} the self-adaptive computing system encounters, or by new goals that are added to the system. When encountering such an event, unanticipated change detection will trigger the evolutionary self-learning engine to start an evolution of the self-adaptive computing systems (see below). 

As an example, assume that anomaly detection (middle box left in Figure~\ref{fig:blueprint-detail}) discovers that the lenses of cameras are dirty resulting in poor quality images. To deal with this problem, a \textit{noise removal learning module} is added to the computing warehouse that offers a new learning algorithm, for instance a convolutional neural network to handle noisy images. This module will then be used by the evolutionary learning engine for evolving the architecture configuration of the self-adaptive computing system (further explained below). 
As another example, consider the introduction of a new goal to reduce pollution in the smart city area caused by mobility. To deal with this new goal, an operator adds a new \textit{pollution reduction goal} to the evolutionary learning engine via the dashboard. This will trigger the evolutionary learning engine to start an evolution of the self-adaptive computing system taking into account this new goal (further explained below). 

\vspace{5pt}\noindent\textbf{Evolutionary Self-learning Engine}. At the heart of a \name{} is an \textit{evolutionary learning engine} that autonomously evolves the self-adaptive computing system to handle any unanticipated changes that cannot be handled by the build-in learning and adaptation mechanisms, realizing requirement (4). When anticipated change detection discovers an anomaly or when a new goal is added to the system, the evolutionary learning engine starts to evolve its internal \textit{model of the self-adaptive computing system}. This runtime model contains an up-to-date representation of the architecture of the self-adaptive computing system along with its goals (self-awareness), and relevant parts of the environment (context-awareness). The evolution of the model is conducted by an \textit{evolutionary learning pipeline} that evolves the architectural configuration of the self-adaptive computing system to obtain its goals.~During this process, the engine may integrate new auto-evolution-enabled computing elements provided by computing warehouses as needed. To evolve the system architecture, the engine runs experiments, executing different subsequent variants of the evolved model in a sandbox.~Using suitable metrics for assessing the performance of the evolving architectural models of the self-adaptive computing system in each evolutionary step, the engine will optimize the self-adaptive computing system model, resulting in a novel architecture that mitigates the unanticipated change that triggered the evolution. During the experiments, the engine may exploit \textit{historical data}, for instance to train a learning module, and experimental results may be stored for reuse later.~\textit{Change enactment} will then replace the running architecture of the self-adaptive computing system with the novel architecture.  

As an example, when discovering that the lenses of cameras are dirty (continuing the example above), the evolutionary learning engine (middle box in Figure~\ref{fig:blueprint-detail}) searches the computing warehouse for a solution. Based on the shared ontology, the engine identifies the new noise removal learning module. The evolutionary learning engine then runs online experiments in the \textit{sandbox}, evolving the model of the current architecture of the self-adaptive computing system and integrating the new noise removal learning module. The engine will use the resolution and quality improvements of images as performance metrics. During this process, the engine may exploit historical data to accelerate the evolution process, and particular experimental results may be stored for later usage. Once the novel architecture is identified that satisfies the system goals, the current configuration will be evolved through change enactment. 

As another example, when the new pollution reduction goal is added to the system (continuing the other example above), the evolutionary learning engine will search in the catalog of the computing warehouse and find the (newly added) pollution sensor module. Based on the usage guidance provided by this module a set of new \textit{pollution sensors} will be activated in the smart city area (possibly involving a field worker). The evolutionary self-learning pipeline will then evolve the architecture of the self-adaptive computing system by extending the mobility tracking platform with functionality to track air pollution and set the configurations of the sensors via the mobility control platform (both derived from the pollution sensor module). Furthermore, the new goal will be added to the adaptation manager. Finally, the learning module will be enhanced to take into account the data of the mobility data module produced by the pollution sensors. To configure the learning model, the engine may use historical data collected by the system. Once the new architecture is configured, it can be deployed via change enactment enabling the smart city mobility system to reduce the pollution by adjusting its settings, e.g. adapting conventional traffic via smart traffic boards. 

\vspace{5pt}\noindent\textbf{Evolution Guidance}. Depending on the domain at hand, human experts may be involved to \textit{guide the evolution} of a \name{}, realizing requirement (5). Evolution guidance can range from a basic dashboard that shows key performance indicators of a \name{} and offers ``knobs'' allowing operators to upload new computing elements, add new goals or define 
constraints on the behavior of the system to ensure its trustworthiness, up to full-fledged embodied AI that exploits intelligent user interfaces enabling operators to guide the evolution process of \name{}s interactively~\citep{KephartDESTD19}.~New goals or constraints may refer to various concerns of users, such as performance, safety, privacy, energy consumption, environmental protection, or ethics.~Evolution guidance may include the option for operators to provide feedback about discovered \mbox{anomalies or give advice on architecture evolution at the evolutionary learning engine, among others.} 

For instance, in the smart city mobility scenario, see  Figure~\ref{fig:blueprint-detail} (box at the top), evolution guidance enables software developers to add new auto-evolution-enabled modules to  the computing warehouse, such as a new learning module for noise removal. Evolution guidance also offers an interactive dashboard enabling an operator to support the evolutionary learning engine with identifying new software architectures of the computing-learning system. For instance, the operator may suggest (possibly new) quantitative and qualitative criteria (goals) to guide a evolutionary pipeline in identifying new architectural configurations. The feedback of the operator may be incorporated into the fitness function allowing the learning pipeline to distinguish between promising and poor architectural configurations when evolving the model of the self-adaptive computing system, enhancing its performance.

\section{Research Challenges Ahead}\label{sec:conclusions}

To conclude, we summarize the novelty of \names{}, highlight key challenges to realize the vision of \name{}s, and provide starting points to tackle them. 

Smart approaches have demonstrated their value for dealing with changes \textit{within the operational domain} of computing system that are composed of regular computing elements. \Names{} extends this to the operational domain of computing systems that integrate regular computing elements with learning algorithms, enabling these systems to deal with a vast amount of highly complex data. Currently, we rely on software evolution to deal with \textit{outside the operational domain}, i.e., changes that were not anticipated when the system was built and deployed. The evolution of software systems is currently still a human-driven process that is supported by tools that automate the continuous integration and deployment of new computing elements. Lifelong learning provides the means to deal with new tasks during operation, yet, this evolution targets learning algorithms. \Names{} on the other hand exploits computing warehouses, enabling \name{}s to evolve autonomously, thereby selecting and integrating new computing elements autonomously during operation based on the needs at hand. Optionally, humans can offer support to \name{}s, for instance, for setting goals on performance, safety, privacy, etc., and providing guidance to support the evolutionary learning process if needed.  

We motivated and described how \names{} enables dealing with the lasting problem of how to engineer long running computing systems that can autonomously adapt and evolve to deal with ever changing conditions, anticipated and unanticipated.~Yet, realizing the vision of \names{}, raises fundamental challenges. We list six key achievements that are required to tackle these challenges: 

\begin{enumerate}
    \item A novel overarching modeling approach for the design of self-evolving computing systems. Contrary to traditional software architecture design languages~\citep{muccini2021software}, a new modeling approach is required that should provide first-class support for specifying heterogenous computing systems that integrate computing and learning elements, as well as the different types of building blocks of self-evolving computing systems. This modeling approach will enable a designer to analyse the compliance of the model of a self-evolving computing system with its high-level goals. 
    
    \item The definition of standardized representations and interfaces of auto-evolution-enabled computing elements (regular and learning elements) that can be seamlessly integrated by self-evolving computing systems. Contrary to existing component-based modeling approaches, see e.g.,~\citep{10.1007/978-3-540-24774-6_3}, auto-evolution-enabled computing elements require two types of meta data: (i) meta data that enables self-evolving computing systems to characterize elements and select an element as needed, and (ii) meta data to incorporate a selected element autonomously. The first type of meta data is similar to a ``data sheet,'' while the second type is similar to a ``usage guide.'' Enabling self-evolving computing systems to reason about and integrate auto-evolution-enabled computing elements require both a well-defined (standardized) syntax and a shared ontology. 

    \item Novel methods and algorithms for realizing self-adaptation of heterogeneous computing systems that need to deal with conflicting goals and operate under uncertainty and resource constraints. An interesting approach to tackle this challenge is the use of dynamic, preference-based, multi-objective, on-line optimization, leveraging state-of-the-art knee-point identification~\citep{8638825}, and preference-based~\citep{10.1145/3205455.3205497}, and on-line optimization~\citep{7886303}. Here the Pareto-frontier becomes a moving target, while the objectives can change when an architecture evolution is applied by the self-evolving computing system.
 
    \item A novel family of anomaly and novelty discovering methods for complex high-dimensional data relying on unsupervised learning. One approach to tackle this challenge is to model the data as a union of low-dimensional manifolds~\citep{you2017provable}. Anomalies are then data points that do not lie in any manifold, i.e., outliers, while novelties are data points that belong to a new manifold, e.g., a new class. The challenge here will be to identify nonlinear manifolds that change over time. Additionally, the solution should be able to deal with multi-modal on-line data streams, e.g., leveraging temporal convolutional autoencoders~\citep{THILL2021107751}. 
    
    \item A novel evolutionary self-learning pipeline for evolving heterogeneous computing systems to deal with unanticipated changes (anomalies, novelties, new goals). Core to such a solution will be: up-to-date architectural models of the underlying heterogenous system with its goals and constraints, and the context in which the system operates. These models should account for the evolution of the system. Evolving the current software architecture requires suitable architectural variation operators that comply with the syntactical and semantical constraints of the evolving architecture. One approach to tackle this is using a (1,$\lambda$) algorithm~\citep{Baeck2013} that selects the best ``offspring'' and iterates the evolution through simulation in a sandbox. New evaluation functions with guarantees will be required, e.g., leveraging statistical model checking of the candidate architectures, and (ii) preference-based, multi-objective optimization providing approximations of Pareto optimality.

    \item Novel notations and mechanisms that enable system operators to add new goals and interact with the evolutionary self-learning pipeline. Changing goals is an important trigger for evolving computing systems. This requires a dashboard for humans to interact with the system and modify its goals. Unlike existing goal models (e.g.,~\cite{cheng09}), self-evolving computing systems require models that evolve dynamically. Goals may be provided with meta data that refers to elements of computing warehouses (e.g., a goal for a new modality of traffic may have meta data about sensors and software to track that traffic). The models should provide mechanisms that automatically translate the changes of the goals to a format that can act as a trigger to evolution. A self-evolving system may be equipped with mechanisms that enable the system to communicate the options for evolution and ask the human to advise on the selection if needed. The dashboard may supports this type of interaction. For instance, the system may show a subset of candidate architectural configurations along with a qualification of the options. The human may then select one of the options to continue the evolution process, leveraging for example~\cite{10.1007/978-3-030-72062-9_22}. 
    
\end{enumerate}

An additional open challenge is how to handle the need for dynamic resource management. While the warehouse may to some degree deal with resource provision and management, the acquisition of hardware and other resources that are needed to support self-evolution may require dedicated support. 

Addressing these challenges requires the combined expertise in a variety of areas: dynamic software architectures and scalable and trustworthy approaches for self-adaptation (to deal with the challenges of adaptation of heterogeneous computing systems), unsupervised learning and runtime goal models (to deal with the challenges of unanticipated change detection), self-awareness, dynamic learning architectures, and evolutionary learning mechanisms (to deal with the challenges of evolutionary learning), and software engineering (to deal with the challenges of computing warehouses and evolution guidance).
Only the synergy between these specializations can adequately yield solutions to realize the vision of \names{}.

\bibliographystyle{ACM-Reference-Format}
\bibliography{main}


\begin{thebibliography}{89}


\ifx \showCODEN    \undefined \def \showCODEN     #1{\unskip}     \fi
\ifx \showDOI      \undefined \def \showDOI       #1{#1}\fi
\ifx \showISBNx    \undefined \def \showISBNx     #1{\unskip}     \fi
\ifx \showISBNxiii \undefined \def \showISBNxiii  #1{\unskip}     \fi
\ifx \showISSN     \undefined \def \showISSN      #1{\unskip}     \fi
\ifx \showLCCN     \undefined \def \showLCCN      #1{\unskip}     \fi
\ifx \shownote     \undefined \def \shownote      #1{#1}          \fi
\ifx \showarticletitle \undefined \def \showarticletitle #1{#1}   \fi
\ifx \showURL      \undefined \def \showURL       {\relax}        \fi
\providecommand\bibfield[2]{#2}
\providecommand\bibinfo[2]{#2}
\providecommand\natexlab[1]{#1}
\providecommand\showeprint[2][]{arXiv:#2}

\bibitem[\protect\citeauthoryear{Andersson, Baresi, Bencomo, de~Lemos, Gorla,
  Inverardi, and Vogel}{Andersson et~al\mbox{.}}{2013}]%
        {Andersson2013}
\bibfield{author}{\bibinfo{person}{J. Andersson}, \bibinfo{person}{L. Baresi},
  \bibinfo{person}{N. Bencomo}, \bibinfo{person}{R. de Lemos},
  \bibinfo{person}{A. Gorla}, \bibinfo{person}{P. Inverardi}, {and}
  \bibinfo{person}{T. Vogel}.} \bibinfo{year}{2013}\natexlab{}.
\newblock \bibinfo{booktitle}{\emph{Software Engineering Processes for
  Self-Adaptive Systems}}.
\newblock \bibinfo{publisher}{Springer}, \bibinfo{pages}{51--75}.
\newblock
\showISBNx{978-3-642-35813-5}
\urldef\tempurl%
\url{https://doi.org/10.1007/978-3-642-35813-5_3}
\showDOI{\tempurl}


\bibitem[\protect\citeauthoryear{B\"ack, Foussette, and Krause}{B\"ack
  et~al\mbox{.}}{2013}]%
        {Baeck2013}
\bibfield{author}{\bibinfo{person}{T. B\"ack}, \bibinfo{person}{C. Foussette},
  {and} \bibinfo{person}{P. Krause}.} \bibinfo{year}{2013}\natexlab{}.
\newblock \bibinfo{booktitle}{\emph{Contemporary Evolution Strategies}}.
\newblock \bibinfo{publisher}{Natural Computing Series, Springer}.
\newblock


\bibitem[\protect\citeauthoryear{Baresi and Ghezzi}{Baresi and Ghezzi}{2010}]%
        {1882362.1882367}
\bibfield{author}{\bibinfo{person}{L. Baresi} {and} \bibinfo{person}{C.
  Ghezzi}.} \bibinfo{year}{2010}\natexlab{}.
\newblock \showarticletitle{The Disappearing Boundary between Development-Time
  and Run-Time}. In \bibinfo{booktitle}{\emph{{Future of Software Engineering
  Research}}}. \bibinfo{publisher}{ACM}, \bibinfo{pages}{17–22}.
\newblock
\showISBNx{9781450304276}
\urldef\tempurl%
\url{https://doi.org/10.1145/1882362.1882367}
\showDOI{\tempurl}


\bibitem[\protect\citeauthoryear{Bennett and Rajlich}{Bennett and
  Rajlich}{2000}]%
        {10.1145-336512.336534}
\bibfield{author}{\bibinfo{person}{K. Bennett} {and} \bibinfo{person}{V.
  Rajlich}.} \bibinfo{year}{2000}\natexlab{}.
\newblock \showarticletitle{Software Maintenance and Evolution: A Roadmap}. In
  \bibinfo{booktitle}{\emph{Conference on The Future of Software Engineering}}
  (Limerick, Ireland) \emph{(\bibinfo{series}{ICSE '00})}.
  \bibinfo{publisher}{Association for Computing Machinery},
  \bibinfo{address}{New York, NY, USA}, \bibinfo{pages}{73–87}.
\newblock
\showISBNx{1581132530}
\urldef\tempurl%
\url{https://doi.org/10.1145/336512.336534}
\showDOI{\tempurl}


\bibitem[\protect\citeauthoryear{Bernardo and Hillston}{Bernardo and
  Hillston}{2007}]%
        {DBLP:conf/sfm/2007}
\bibfield{editor}{\bibinfo{person}{M. Bernardo} {and} \bibinfo{person}{J.
  Hillston}} (Eds.). \bibinfo{year}{2007}\natexlab{}.
\newblock \bibinfo{booktitle}{\emph{Formal Methods for Performance Evaluation,
  7th International School on Formal Methods for the Design of Computer,
  Communication, and Software Systems, {SFM} 2007, Bertinoro, Italy, May
  28-June 2, 2007, Advanced Lectures}}. \bibinfo{series}{Lecture Notes in
  Computer Science}, Vol.~\bibinfo{volume}{4486}.
  \bibinfo{publisher}{Springer}.
\newblock


\bibitem[\protect\citeauthoryear{{Blair}, {Bencomo}, and {France}}{{Blair}
  et~al\mbox{.}}{2009}]%
        {Blair2009}
\bibfield{author}{\bibinfo{person}{G. {Blair}}, \bibinfo{person}{N. {Bencomo}},
  {and} \bibinfo{person}{R.~B. {France}}.} \bibinfo{year}{2009}\natexlab{}.
\newblock \showarticletitle{Models@ run.time}.
\newblock \bibinfo{journal}{\emph{Computer}} \bibinfo{volume}{42},
  \bibinfo{number}{10} (\bibinfo{year}{2009}), \bibinfo{pages}{22--27}.
\newblock
\urldef\tempurl%
\url{https://doi.org/10.1109/MC.2009.326}
\showDOI{\tempurl}


\bibitem[\protect\citeauthoryear{Boukerche, Zheng, and Alfandi}{Boukerche
  et~al\mbox{.}}{2020}]%
        {3381028}
\bibfield{author}{\bibinfo{person}{A. Boukerche}, \bibinfo{person}{L. Zheng},
  {and} \bibinfo{person}{O. Alfandi}.} \bibinfo{year}{2020}\natexlab{}.
\newblock \showarticletitle{Outlier Detection: Methods, Models, and
  Classification}.
\newblock \bibinfo{journal}{\emph{ACM Comput. Surv.}} \bibinfo{volume}{53},
  \bibinfo{number}{3}, Article \bibinfo{articleno}{55} (\bibinfo{date}{June}
  \bibinfo{year}{2020}), \bibinfo{numpages}{37}~pages.
\newblock
\showISSN{0360-0300}
\urldef\tempurl%
\url{https://doi.org/10.1145/3381028}
\showDOI{\tempurl}


\bibitem[\protect\citeauthoryear{Bruneton, Coupaye, Leclercq, Quema, and
  Stefani}{Bruneton et~al\mbox{.}}{2004}]%
        {10.1007/978-3-540-24774-6_3}
\bibfield{author}{\bibinfo{person}{E. Bruneton}, \bibinfo{person}{T. Coupaye},
  \bibinfo{person}{M. Leclercq}, \bibinfo{person}{V. Quema}, {and}
  \bibinfo{person}{J-B. Stefani}.} \bibinfo{year}{2004}\natexlab{}.
\newblock \showarticletitle{An Open Component Model and Its Support in Java}.
  In \bibinfo{booktitle}{\emph{Component-Based Software Engineering}}.
  \bibinfo{publisher}{Springer}, \bibinfo{pages}{7--22}.
\newblock


\bibitem[\protect\citeauthoryear{Buckley, Mens, Zenger, Rashid, and
  Kniesel}{Buckley et~al\mbox{.}}{2005}]%
        {10.5555-1090744.1090746}
\bibfield{author}{\bibinfo{person}{J. Buckley}, \bibinfo{person}{T. Mens},
  \bibinfo{person}{M. Zenger}, \bibinfo{person}{A. Rashid}, {and}
  \bibinfo{person}{G.r Kniesel}.} \bibinfo{year}{2005}\natexlab{}.
\newblock \showarticletitle{Towards a Taxonomy of Software Change: Research
  Articles}.
\newblock \bibinfo{journal}{\emph{Journal on Software Maintenance and
  Evolution}} \bibinfo{volume}{17}, \bibinfo{number}{5} (\bibinfo{date}{Sept.}
  \bibinfo{year}{2005}), \bibinfo{pages}{309–332}.
\newblock
\showISSN{1532-060X}


\bibitem[\protect\citeauthoryear{Bures, Weyns, Schmerl, Tovar, Boden, Gabor,
  Gerostathopoulos, Gupta, Kang, Knauss, Patel, Rashid, Ruchkin, Sukkerd, and
  Tsigkanos}{Bures et~al\mbox{.}}{2017}]%
        {3089649.3089656}
\bibfield{author}{\bibinfo{person}{T. Bures}, \bibinfo{person}{D. Weyns},
  \bibinfo{person}{B. Schmerl}, \bibinfo{person}{E. Tovar}, \bibinfo{person}{E.
  Boden}, \bibinfo{person}{T. Gabor}, \bibinfo{person}{I. Gerostathopoulos},
  \bibinfo{person}{P. Gupta}, \bibinfo{person}{E. Kang}, \bibinfo{person}{A.
  Knauss}, \bibinfo{person}{P. Patel}, \bibinfo{person}{A. Rashid},
  \bibinfo{person}{I. Ruchkin}, \bibinfo{person}{R. Sukkerd}, {and}
  \bibinfo{person}{C. Tsigkanos}.} \bibinfo{year}{2017}\natexlab{}.
\newblock \showarticletitle{Software Engineering for Smart Cyber-Physical
  Systems: Challenges and Promising Solutions}.
\newblock \bibinfo{journal}{\emph{SIGSOFT Software Engineering Notes}}
  \bibinfo{volume}{42}, \bibinfo{number}{2} (\bibinfo{year}{2017}),
  \bibinfo{pages}{19–24}.
\newblock
\showISSN{0163-5948}
\urldef\tempurl%
\url{https://doi.org/10.1145/3089649.3089656}
\showDOI{\tempurl}


\bibitem[\protect\citeauthoryear{{Calinescu}, {Mirandola}, {Perez-Palacin}, and
  {Weyns}}{{Calinescu} et~al\mbox{.}}{2020}]%
        {9196226}
\bibfield{author}{\bibinfo{person}{R. {Calinescu}}, \bibinfo{person}{R.
  {Mirandola}}, \bibinfo{person}{D. {Perez-Palacin}}, {and} \bibinfo{person}{D.
  {Weyns}}.} \bibinfo{year}{2020}\natexlab{}.
\newblock \showarticletitle{Understanding Uncertainty in Self-adaptive
  Systems}. In \bibinfo{booktitle}{\emph{IEEE International Conference on
  Autonomic Computing and Self-Organizing Systems}}. \bibinfo{pages}{242--251}.
\newblock
\urldef\tempurl%
\url{https://doi.org/10.1109/ACSOS49614.2020.00047}
\showDOI{\tempurl}


\bibitem[\protect\citeauthoryear{Calinescu, Weyns, Gerasimou, Iftikhar, Habli,
  and Kelly}{Calinescu et~al\mbox{.}}{2018}]%
        {8008800}
\bibfield{author}{\bibinfo{person}{R. Calinescu}, \bibinfo{person}{D. Weyns},
  \bibinfo{person}{S. Gerasimou}, \bibinfo{person}{M.~U. Iftikhar},
  \bibinfo{person}{I. Habli}, {and} \bibinfo{person}{T. Kelly}.}
  \bibinfo{year}{2018}\natexlab{}.
\newblock \showarticletitle{Engineering Trustworthy Self-Adaptive Software with
  Dynamic Assurance Cases}.
\newblock \bibinfo{journal}{\emph{IEEE Transactions on Software Engineering}}
  \bibinfo{volume}{44}, \bibinfo{number}{11} (\bibinfo{year}{2018}),
  \bibinfo{pages}{1039--1069}.
\newblock
\urldef\tempurl%
\url{https://doi.org/10.1109/TSE.2017.2738640}
\showDOI{\tempurl}


\bibitem[\protect\citeauthoryear{Camazine, Deneubourg, Franks, Sneyd,
  Theraulas, and Bonabeau}{Camazine et~al\mbox{.}}{2003}]%
        {Camazine:2003}
\bibfield{author}{\bibinfo{person}{S. Camazine}, \bibinfo{person}{J-L.
  Deneubourg}, \bibinfo{person}{N. Franks}, \bibinfo{person}{J. Sneyd},
  \bibinfo{person}{G. Theraulas}, {and} \bibinfo{person}{E. Bonabeau}.}
  \bibinfo{year}{2003}\natexlab{}.
\newblock \bibinfo{booktitle}{\emph{Organization in Biological Systems}}.
\newblock \bibinfo{publisher}{Princeton Studies in Complexity},
  \bibinfo{address}{USA}.
\newblock
\showISBNx{9780691116242}


\bibitem[\protect\citeauthoryear{Chapin, Hale, Kham, Ramil, and Tan}{Chapin
  et~al\mbox{.}}{2001}]%
        {10.5555-371697.371701}
\bibfield{author}{\bibinfo{person}{N. Chapin}, \bibinfo{person}{J. Hale},
  \bibinfo{person}{K. Kham}, \bibinfo{person}{J. Ramil}, {and}
  \bibinfo{person}{W. Tan}.} \bibinfo{year}{2001}\natexlab{}.
\newblock \showarticletitle{Types of Software Evolution and Software
  Maintenance}.
\newblock \bibinfo{journal}{\emph{Journal of Software Maintenance}}
  \bibinfo{volume}{13}, \bibinfo{number}{1} (\bibinfo{year}{2001}),
  \bibinfo{pages}{3–30}.
\newblock
\showISSN{1040-550X}


\bibitem[\protect\citeauthoryear{Chen, Tino, Rodan, and Yao}{Chen
  et~al\mbox{.}}{2014}]%
        {6515601}
\bibfield{author}{\bibinfo{person}{H. Chen}, \bibinfo{person}{P. Tino},
  \bibinfo{person}{A. Rodan}, {and} \bibinfo{person}{X. Yao}.}
  \bibinfo{year}{2014}\natexlab{}.
\newblock \showarticletitle{Learning in the Model Space for Cognitive Fault
  Diagnosis}.
\newblock \bibinfo{journal}{\emph{IEEE Transactions on Neural Networks and
  Learning Systems}} \bibinfo{volume}{25}, \bibinfo{number}{1}
  (\bibinfo{year}{2014}), \bibinfo{pages}{124--136}.
\newblock
\urldef\tempurl%
\url{https://doi.org/10.1109/TNNLS.2013.2256797}
\showDOI{\tempurl}


\bibitem[\protect\citeauthoryear{Chen, Li, and Yao}{Chen
  et~al\mbox{.}}{2018b}]%
        {7886303}
\bibfield{author}{\bibinfo{person}{R. Chen}, \bibinfo{person}{K. Li}, {and}
  \bibinfo{person}{X. Yao}.} \bibinfo{year}{2018}\natexlab{b}.
\newblock \showarticletitle{Dynamic Multiobjectives Optimization With a
  Changing Number of Objectives}.
\newblock \bibinfo{journal}{\emph{IEEE Transactions on Evolutionary
  Computation}} \bibinfo{volume}{22}, \bibinfo{number}{1}
  (\bibinfo{year}{2018}), \bibinfo{pages}{157--171}.
\newblock
\urldef\tempurl%
\url{https://doi.org/10.1109/TEVC.2017.2669638}
\showDOI{\tempurl}


\bibitem[\protect\citeauthoryear{Chen, Bahsoon, and Yao}{Chen
  et~al\mbox{.}}{2018a}]%
        {10.1145-3190507}
\bibfield{author}{\bibinfo{person}{T. Chen}, \bibinfo{person}{R. Bahsoon},
  {and} \bibinfo{person}{X. Yao}.} \bibinfo{year}{2018}\natexlab{a}.
\newblock \showarticletitle{A Survey and Taxonomy of Self-Aware and
  Self-Adaptive Cloud Autoscaling Systems}.
\newblock \bibinfo{journal}{\emph{ACM Comput. Surv.}} \bibinfo{volume}{51},
  \bibinfo{number}{3}, Article \bibinfo{articleno}{61} (\bibinfo{date}{June}
  \bibinfo{year}{2018}), \bibinfo{numpages}{40}~pages.
\newblock
\showISSN{0360-0300}
\urldef\tempurl%
\url{https://doi.org/10.1145/3190507}
\showDOI{\tempurl}


\bibitem[\protect\citeauthoryear{Chen, Bahsoon, and Yao}{Chen
  et~al\mbox{.}}{2020}]%
        {9089006}
\bibfield{author}{\bibinfo{person}{T. Chen}, \bibinfo{person}{R. Bahsoon},
  {and} \bibinfo{person}{X. Yao}.} \bibinfo{year}{2020}\natexlab{}.
\newblock \showarticletitle{Synergizing Domain Expertise With Self-Awareness in
  Software Systems: A Patternized Architecture Guideline}.
\newblock \bibinfo{journal}{\emph{Proc. IEEE}} \bibinfo{volume}{108},
  \bibinfo{number}{7} (\bibinfo{year}{2020}), \bibinfo{pages}{1094--1126}.
\newblock
\urldef\tempurl%
\url{https://doi.org/10.1109/JPROC.2020.2985293}
\showDOI{\tempurl}


\bibitem[\protect\citeauthoryear{Chen, Li, Bahsoon, and Yao}{Chen
  et~al\mbox{.}}{2018c}]%
        {10.1145-3204459}
\bibfield{author}{\bibinfo{person}{T. Chen}, \bibinfo{person}{K. Li},
  \bibinfo{person}{R. Bahsoon}, {and} \bibinfo{person}{X. Yao}.}
  \bibinfo{year}{2018}\natexlab{c}.
\newblock \showarticletitle{FEMOSAA: Feature-Guided and Knee-Driven
  Multi-Objective Optimization for Self-Adaptive Software}.
\newblock \bibinfo{journal}{\emph{ACM Transactions on Software Engineering and
  Methodology}} \bibinfo{volume}{27}, \bibinfo{number}{2}, Article
  \bibinfo{articleno}{5} (\bibinfo{year}{2018}), \bibinfo{numpages}{50}~pages.
\newblock
\showISSN{1049-331X}
\urldef\tempurl%
\url{https://doi.org/10.1145/3204459}
\showDOI{\tempurl}


\bibitem[\protect\citeauthoryear{Chen and Liu}{Chen and Liu}{2018}]%
        {LML}
\bibfield{author}{\bibinfo{person}{Z. Chen} {and} \bibinfo{person}{B. Liu}.}
  \bibinfo{year}{2018}\natexlab{}.
\newblock \bibinfo{booktitle}{\emph{Lifelong Machine Learning}}.
\newblock \bibinfo{publisher}{Morgan \& Claypool}.
\newblock
\showISBNx{9781681733029}


\bibitem[\protect\citeauthoryear{Cheng et~al\mbox{.}}{Cheng
  et~al\mbox{.}}{2009a}]%
        {Cheng2009}
\bibfield{author}{\bibinfo{person}{B. Cheng} {et~al\mbox{.}}}
  \bibinfo{year}{2009}\natexlab{a}.
\newblock \bibinfo{booktitle}{\emph{Software Engineering for Self-Adaptive
  Systems: A Research Roadmap}}.
\newblock \bibinfo{publisher}{Springer}, \bibinfo{pages}{1--26}.
\newblock
\showISBNx{978-3-642-02161-9}
\urldef\tempurl%
\url{https://doi.org/10.1007/978-3-642-02161-9_1}
\showDOI{\tempurl}


\bibitem[\protect\citeauthoryear{Cheng, Sawyer, Bencomo, and Whittle}{Cheng
  et~al\mbox{.}}{2009b}]%
        {cheng09}
\bibfield{author}{\bibinfo{person}{B. Cheng}, \bibinfo{person}{P. Sawyer},
  \bibinfo{person}{N. Bencomo}, {and} \bibinfo{person}{J. Whittle}.}
  \bibinfo{year}{2009}\natexlab{b}.
\newblock \showarticletitle{A Goal-Based Modeling Approach to Develop
  Requirements of an Adaptive System with Environmental Uncertainty}. In
  \bibinfo{booktitle}{\emph{Model Driven Engineering Languages and Systems}}.
  \bibinfo{publisher}{Springer}.
\newblock


\bibitem[\protect\citeauthoryear{Dastani and Testerink}{Dastani and
  Testerink}{2016}]%
        {Dastani16}
\bibfield{author}{\bibinfo{person}{M. Dastani} {and} \bibinfo{person}{B.
  Testerink}.} \bibinfo{year}{2016}\natexlab{}.
\newblock \showarticletitle{Design patterns for multi-agent programming}.
\newblock \bibinfo{journal}{\emph{International Journal of Agent-Oriented
  Software Engineering}} \bibinfo{volume}{5}, \bibinfo{number}{2-3}
  (\bibinfo{year}{2016}), \bibinfo{pages}{167--202}.
\newblock


\bibitem[\protect\citeauthoryear{de~Winter, van Stein, and B{\"a}ck}{de~Winter
  et~al\mbox{.}}{2021}]%
        {10.1007/978-3-030-72062-9_22}
\bibfield{author}{\bibinfo{person}{R. de Winter}, \bibinfo{person}{B. van
  Stein}, {and} \bibinfo{person}{T. B{\"a}ck}.}
  \bibinfo{year}{2021}\natexlab{}.
\newblock \showarticletitle{SAMO-COBRA: A Fast Surrogate Assisted Constrained
  Multi-objective Optimization Algorithm}. In
  \bibinfo{booktitle}{\emph{Evolutionary Multi-Criterion Optimization}}.
  \bibinfo{publisher}{Springer}.
\newblock


\bibitem[\protect\citeauthoryear{Dearle}{Dearle}{2007}]%
        {10.1109-FOSE.2007.20}
\bibfield{author}{\bibinfo{person}{A. Dearle}.}
  \bibinfo{year}{2007}\natexlab{}.
\newblock \showarticletitle{Software Deployment, Past, Present and Future}. In
  \bibinfo{booktitle}{\emph{2007 Future of Software Engineering}}.
  \bibinfo{publisher}{IEEE Computer Society}, \bibinfo{address}{USA},
  \bibinfo{pages}{269–284}.
\newblock
\showISBNx{0769528295}
\urldef\tempurl%
\url{https://doi.org/10.1109/FOSE.2007.20}
\showDOI{\tempurl}


\bibitem[\protect\citeauthoryear{Denkena and Morke}{Denkena and Morke}{2017}]%
        {Denkena:2017}
\bibfield{author}{\bibinfo{person}{B. Denkena} {and} \bibinfo{person}{T.
  Morke}.} \bibinfo{year}{2017}\natexlab{}.
\newblock \bibinfo{booktitle}{\emph{Cyber-Physical and Gentelligent Systems in
  Manufacturing and Life Cycle}}.
\newblock \bibinfo{publisher}{Academic Press}.
\newblock
\showISBNx{978-0-12-811939-6}


\bibitem[\protect\citeauthoryear{Det-Norske-Veritas}{Det-Norske-Veritas}{2020}]%
        {DNV}
\bibfield{author}{\bibinfo{person}{Det-Norske-Veritas}.}
  \bibinfo{year}{2020}\natexlab{}.
\newblock \showarticletitle{Technology Outlook 2030 - Safer, Smarter, Greener}.
\newblock  (\bibinfo{year}{2020}), \bibinfo{pages}{1--110}.
\newblock
\urldef\tempurl%
\url{www.dnvgl.com}
\showURL{%
\tempurl}


\bibitem[\protect\citeauthoryear{Dick and Naumann}{Dick and Naumann}{2010}]%
        {Dick2010}
\bibfield{author}{\bibinfo{person}{M. Dick} {and} \bibinfo{person}{S.
  Naumann}.} \bibinfo{year}{2010}\natexlab{}.
\newblock \showarticletitle{Enhancing Software Engineering Processes towards
  Sustainable Software Product Design}. In
  \bibinfo{booktitle}{\emph{Integration of Environmental Information in
  Europe}}, \bibfield{editor}{\bibinfo{person}{Klaus Greve} {and}
  \bibinfo{person}{Armin~B. Cremers}} (Eds.). \bibinfo{publisher}{Shaker
  Verlag}, \bibinfo{address}{Aachen}.
\newblock


\bibitem[\protect\citeauthoryear{Dingsøyr, Nerur, Balijepally, and
  Moe}{Dingsøyr et~al\mbox{.}}{2012}]%
        {DINGSOYR20121213}
\bibfield{author}{\bibinfo{person}{T. Dingsøyr}, \bibinfo{person}{S. Nerur},
  \bibinfo{person}{V. Balijepally}, {and} \bibinfo{person}{N. Moe}.}
  \bibinfo{year}{2012}\natexlab{}.
\newblock \showarticletitle{A decade of agile methodologies: Towards explaining
  agile software development}.
\newblock \bibinfo{journal}{\emph{Journal of Systems and Software}}
  \bibinfo{volume}{85}, \bibinfo{number}{6} (\bibinfo{year}{2012}),
  \bibinfo{pages}{1213--1221}.
\newblock
\showISSN{0164-1212}
\urldef\tempurl%
\url{https://doi.org/10.1016/j.jss.2012.02.033}
\showDOI{\tempurl}
\newblock
\shownote{Special Issue: Agile Development.}


\bibitem[\protect\citeauthoryear{Elhabbash, Salama, Bahsoon, and
  Tino}{Elhabbash et~al\mbox{.}}{2019}]%
        {10.1145-3347269}
\bibfield{author}{\bibinfo{person}{A. Elhabbash}, \bibinfo{person}{M. Salama},
  \bibinfo{person}{R. Bahsoon}, {and} \bibinfo{person}{P. Tino}.}
  \bibinfo{year}{2019}\natexlab{}.
\newblock \showarticletitle{Self-Awareness in Software Engineering: A
  Systematic Literature Review}.
\newblock \bibinfo{journal}{\emph{ACM Transactions on Autonomous and Adaptive
  Systems}} \bibinfo{volume}{14}, \bibinfo{number}{2}, Article
  \bibinfo{articleno}{5} (\bibinfo{date}{Oct.} \bibinfo{year}{2019}),
  \bibinfo{numpages}{42}~pages.
\newblock
\showISSN{1556-4665}
\urldef\tempurl%
\url{https://doi.org/10.1145/3347269}
\showDOI{\tempurl}


\bibitem[\protect\citeauthoryear{Esfahani and Malek}{Esfahani and
  Malek}{2013}]%
        {Esfahani2013}
\bibfield{author}{\bibinfo{person}{N. Esfahani} {and} \bibinfo{person}{S.
  Malek}.} \bibinfo{year}{2013}\natexlab{}.
\newblock \bibinfo{booktitle}{\emph{Uncertainty in Self-Adaptive Software
  Systems}}.
\newblock \bibinfo{publisher}{Springer}, \bibinfo{pages}{214--238}.
\newblock
\urldef\tempurl%
\url{https://doi.org/10.1007/978-3-642-35813-5_9}
\showDOI{\tempurl}


\bibitem[\protect\citeauthoryear{European-Commission}{European-Commission}{2021}]%
        {EU}
\bibfield{author}{\bibinfo{person}{European-Commission}.}
  \bibinfo{year}{8/2021}\natexlab{}.
\newblock \showarticletitle{Advanced Computing}.
\newblock  (\bibinfo{year}{8/2021}).
\newblock
\urldef\tempurl%
\url{https://www.nsf.gov/funding/pgm_summ.jsp?pims_id=503306}
\showURL{%
\tempurl}


\bibitem[\protect\citeauthoryear{Garlan, Cheng, Huang, Schmerl, and
  Steenkiste}{Garlan et~al\mbox{.}}{2004}]%
        {Rainbow}
\bibfield{author}{\bibinfo{person}{D. Garlan}, \bibinfo{person}{S. Cheng},
  \bibinfo{person}{A. Huang}, \bibinfo{person}{B. Schmerl}, {and}
  \bibinfo{person}{P. Steenkiste}.} \bibinfo{year}{2004}\natexlab{}.
\newblock \showarticletitle{Rainbow: Architecture-Based Self-Adaptation with
  Reusable Infrastructure}.
\newblock \bibinfo{journal}{\emph{Computer}} \bibinfo{volume}{37},
  \bibinfo{number}{10} (\bibinfo{date}{Oct.} \bibinfo{year}{2004}),
  \bibinfo{pages}{46–54}.
\newblock
\showISSN{0018-9162}
\urldef\tempurl%
\url{https://doi.org/10.1109/MC.2004.175}
\showDOI{\tempurl}


\bibitem[\protect\citeauthoryear{Georgiou, Rizou, and Spinellis}{Georgiou
  et~al\mbox{.}}{2019}]%
        {10.1145/3337773}
\bibfield{author}{\bibinfo{person}{S. Georgiou}, \bibinfo{person}{S. Rizou},
  {and} \bibinfo{person}{D. Spinellis}.} \bibinfo{year}{2019}\natexlab{}.
\newblock \showarticletitle{Software Development Lifecycle for Energy
  Efficiency: Techniques and Tools}.
\newblock \bibinfo{journal}{\emph{ACM Comput. Surv.}} \bibinfo{volume}{52},
  \bibinfo{number}{4}, Article \bibinfo{articleno}{81} (\bibinfo{date}{Aug.}
  \bibinfo{year}{2019}), \bibinfo{numpages}{33}~pages.
\newblock
\showISSN{0360-0300}
\urldef\tempurl%
\url{https://doi.org/10.1145/3337773}
\showDOI{\tempurl}


\bibitem[\protect\citeauthoryear{Gr\"otker}{Gr\"otker}{2015}]%
        {acatech2015}
\bibfield{author}{\bibinfo{person}{R. Gr\"otker}.}
  \bibinfo{year}{2015}\natexlab{}.
\newblock \showarticletitle{New autoMobility: The Future World of Automated
  Road Traffic}.
\newblock \bibinfo{journal}{\emph{National Academy of Science and Engineering,
  acatech Germany}} (\bibinfo{year}{2015}).
\newblock
\showISSN{2192-6166}
\urldef\tempurl%
\url{https://elib.dlr.de/101368/2/acatech_POSITION_PAPER_New_autoMobility_web.pdf}
\showURL{%
\tempurl}


\bibitem[\protect\citeauthoryear{Grubbs}{Grubbs}{1969}]%
        {1969Grubbs}
\bibfield{author}{\bibinfo{person}{F.~E. Grubbs}.}
  \bibinfo{year}{1969}\natexlab{}.
\newblock \showarticletitle{Procedures for detecting outlying observations in
  samples}.
\newblock \bibinfo{journal}{\emph{Technometrics}} \bibinfo{volume}{11},
  \bibinfo{number}{1} (\bibinfo{year}{1969}).
\newblock


\bibitem[\protect\citeauthoryear{Hasselmo}{Hasselmo}{2017}]%
        {HASSELMO2017407}
\bibfield{author}{\bibinfo{person}{M.~E. Hasselmo}.}
  \bibinfo{year}{2017}\natexlab{}.
\newblock \showarticletitle{Avoiding Catastrophic Forgetting}.
\newblock \bibinfo{journal}{\emph{Trends in Cognitive Sciences}}
  \bibinfo{volume}{21}, \bibinfo{number}{6} (\bibinfo{year}{2017}),
  \bibinfo{pages}{407--408}.
\newblock
\showISSN{1364-6613}
\urldef\tempurl%
\url{https://doi.org/10.1016/j.tics.2017.04.001}
\showDOI{\tempurl}


\bibitem[\protect\citeauthoryear{IBM}{IBM}{2003}]%
        {2003IBM}
\bibfield{author}{\bibinfo{person}{IBM}.} \bibinfo{year}{2003}\natexlab{}.
\newblock \showarticletitle{An Architectural Blueprint for Autonomic
  Computing}.
\newblock  (\bibinfo{year}{2003}).
\newblock
\urldef\tempurl%
\url{citeseerx.ist.psu.edu/viewdoc/download?doi=10.1.1.150.1011&rep=rep1&type=pdf}
\showURL{%
\tempurl}


\bibitem[\protect\citeauthoryear{Jackson}{Jackson}{1997}]%
        {1997Jackson}
\bibfield{author}{\bibinfo{person}{M. Jackson}.}
  \bibinfo{year}{1997}\natexlab{}.
\newblock \showarticletitle{The Meaning of Requirements}.
\newblock \bibinfo{journal}{\emph{Annals of Software Engineering. Springer
  10480.}} \bibinfo{volume}{3}, \bibinfo{number}{1} (\bibinfo{year}{1997}),
  \bibinfo{pages}{5--21}.
\newblock
\showISSN{1573-7489}
\urldef\tempurl%
\url{https://doi.org/10.1023/A:1018990005598}
\showDOI{\tempurl}


\bibitem[\protect\citeauthoryear{J{\"a}rvinen, Huomo, Mikkonen, and
  Tyrv{\"a}inen}{J{\"a}rvinen et~al\mbox{.}}{2014}]%
        {978-3-319-08738}
\bibfield{author}{\bibinfo{person}{J. J{\"a}rvinen}, \bibinfo{person}{T.
  Huomo}, \bibinfo{person}{T. Mikkonen}, {and} \bibinfo{person}{P.
  Tyrv{\"a}inen}.} \bibinfo{year}{2014}\natexlab{}.
\newblock \showarticletitle{From Agile Software Development to Mercury
  Business}. In \bibinfo{booktitle}{\emph{Software Business. Towards Continuous
  Value Delivery}}. \bibinfo{publisher}{Springer}.
\newblock


\bibitem[\protect\citeauthoryear{{Jazdi}}{{Jazdi}}{2014}]%
        {6857843}
\bibfield{author}{\bibinfo{person}{N. {Jazdi}}.}
  \bibinfo{year}{2014}\natexlab{}.
\newblock \showarticletitle{Cyber physical systems in the context of Industry
  4.0}. In \bibinfo{booktitle}{\emph{{IEEE International Conference on
  Automation, Quality and Testing, Robotics}}}. \bibinfo{pages}{1--4}.
\newblock
\urldef\tempurl%
\url{https://doi.org/10.1109/AQTR.2014.6857843}
\showDOI{\tempurl}


\bibitem[\protect\citeauthoryear{Juziuk, Weyns, and Holvoet}{Juziuk
  et~al\mbox{.}}{2014}]%
        {JuziukJoanna2014DPfM}
\bibfield{author}{\bibinfo{person}{J. Juziuk}, \bibinfo{person}{D. Weyns},
  {and} \bibinfo{person}{T. Holvoet}.} \bibinfo{year}{2014}\natexlab{}.
\newblock \showarticletitle{Design Patterns for Multi-agent Systems: A
  Systematic Literature Review}.
\newblock In \bibinfo{booktitle}{\emph{Agent-Oriented Software Engineering}}.
  Vol.~\bibinfo{volume}{9783642544323}. \bibinfo{publisher}{Springer},
  \bibinfo{pages}{77--97}.
\newblock
\showISBNx{3642544312}


\bibitem[\protect\citeauthoryear{{Kephart} and {Chess}}{{Kephart} and
  {Chess}}{2003}]%
        {Kephart}
\bibfield{author}{\bibinfo{person}{J. {Kephart}} {and} \bibinfo{person}{D.
  {Chess}}.} \bibinfo{year}{2003}\natexlab{}.
\newblock \showarticletitle{The vision of autonomic computing}.
\newblock \bibinfo{journal}{\emph{Computer}} \bibinfo{volume}{36},
  \bibinfo{number}{1} (\bibinfo{year}{2003}), \bibinfo{pages}{41--50}.
\newblock


\bibitem[\protect\citeauthoryear{Kephart, Dibia, Ellis, Srivastava,
  Talamadupula, and Dholakia}{Kephart et~al\mbox{.}}{2019}]%
        {KephartDESTD19}
\bibfield{author}{\bibinfo{person}{J. Kephart}, \bibinfo{person}{V. Dibia},
  \bibinfo{person}{J. Ellis}, \bibinfo{person}{B. Srivastava},
  \bibinfo{person}{K. Talamadupula}, {and} \bibinfo{person}{M. Dholakia}.}
  \bibinfo{year}{2019}\natexlab{}.
\newblock \showarticletitle{An Embodied Cognitive Assistant for Visualizing and
  Analyzing Exoplanet Data}.
\newblock \bibinfo{journal}{\emph{{IEEE} Internet Computing}}
  \bibinfo{volume}{23}, \bibinfo{number}{2} (\bibinfo{year}{2019}),
  \bibinfo{pages}{31--39}.
\newblock
\urldef\tempurl%
\url{https://doi.org/10.1109/MIC.2019.2906528}
\showDOI{\tempurl}


\bibitem[\protect\citeauthoryear{{Koutsoukos}, {Karsai}, {Laszka}, {Neema},
  {Potteiger}, {Volgyesi}, {Vorobeychik}, and {Sztipanovits}}{{Koutsoukos}
  et~al\mbox{.}}{2018}]%
        {8010538}
\bibfield{author}{\bibinfo{person}{X. {Koutsoukos}}, \bibinfo{person}{G.
  {Karsai}}, \bibinfo{person}{A. {Laszka}}, \bibinfo{person}{H. {Neema}},
  \bibinfo{person}{B. {Potteiger}}, \bibinfo{person}{P. {Volgyesi}},
  \bibinfo{person}{Y. {Vorobeychik}}, {and} \bibinfo{person}{J.
  {Sztipanovits}}.} \bibinfo{year}{2018}\natexlab{}.
\newblock \showarticletitle{SURE: A Modeling and Simulation Integration
  Platform for Evaluation of Secure and Resilient Cyber–Physical Systems}.
\newblock \bibinfo{journal}{\emph{Proc. IEEE}} \bibinfo{volume}{106},
  \bibinfo{number}{1} (\bibinfo{year}{2018}), \bibinfo{pages}{93--112}.
\newblock
\urldef\tempurl%
\url{https://doi.org/10.1109/JPROC.2017.2731741}
\showDOI{\tempurl}


\bibitem[\protect\citeauthoryear{Kramer and Magee}{Kramer and Magee}{2007}]%
        {2007Kramer}
\bibfield{author}{\bibinfo{person}{J. Kramer} {and} \bibinfo{person}{J.
  Magee}.} \bibinfo{year}{2007}\natexlab{}.
\newblock \showarticletitle{Self-Managed Systems: {An} Architectural
  Challenge}. In \bibinfo{booktitle}{\emph{{Future of Software Engineering}}}.
  \bibinfo{publisher}{IEEE}, \bibinfo{pages}{259--268}.
\newblock
\urldef\tempurl%
\url{https://doi.org/10.1109/FOSE.2007.19}
\showDOI{\tempurl}


\bibitem[\protect\citeauthoryear{Lehman and Ramil}{Lehman and Ramil}{2003}]%
        {LEHMAN200333}
\bibfield{author}{\bibinfo{person}{M. Lehman} {and} \bibinfo{person}{J.
  Ramil}.} \bibinfo{year}{2003}\natexlab{}.
\newblock \showarticletitle{Software evolution—Background, theory, practice}.
\newblock \bibinfo{journal}{\emph{Inform. Process. Lett.}}
  \bibinfo{volume}{88}, \bibinfo{number}{1} (\bibinfo{year}{2003}),
  \bibinfo{pages}{33--44}.
\newblock
\showISSN{0020-0190}
\urldef\tempurl%
\url{https://doi.org/10.1016/S0020-0190(03)00382-X}
\showDOI{\tempurl}
\newblock
\shownote{To honour Professor W.M. Turski's Contribution to Computing Science
  on the Occasion of his 65th Birthday.}


\bibitem[\protect\citeauthoryear{Li, Avgeriou, and Liang}{Li
  et~al\mbox{.}}{2015}]%
        {LI2015193}
\bibfield{author}{\bibinfo{person}{Z. Li}, \bibinfo{person}{P. Avgeriou}, {and}
  \bibinfo{person}{P. Liang}.} \bibinfo{year}{2015}\natexlab{}.
\newblock \showarticletitle{A systematic mapping study on technical debt and
  its management}.
\newblock \bibinfo{journal}{\emph{Journal of Systems and Software}}
  \bibinfo{volume}{101} (\bibinfo{year}{2015}), \bibinfo{pages}{193--220}.
\newblock
\showISSN{0164-1212}
\urldef\tempurl%
\url{https://doi.org/10.1016/j.jss.2014.12.027}
\showDOI{\tempurl}


\bibitem[\protect\citeauthoryear{Mahdavi-Hezavehi, Avgeriou, and
  Weyns}{Mahdavi-Hezavehi et~al\mbox{.}}{2017}]%
        {MAHDAVIHEZAVEHI201745}
\bibfield{author}{\bibinfo{person}{S. Mahdavi-Hezavehi}, \bibinfo{person}{P.
  Avgeriou}, {and} \bibinfo{person}{D. Weyns}.}
  \bibinfo{year}{2017}\natexlab{}.
\newblock \showarticletitle{A Classification Framework of Uncertainty in
  Architecture-Based Self-Adaptive Systems With Multiple Quality Requirements}.
\newblock In \bibinfo{booktitle}{\emph{Managing Trade-Offs in Adaptable
  Software Architectures}}, \bibfield{editor}{\bibinfo{person}{I.~Mistrik},
  \bibinfo{person}{N.~Ali}, \bibinfo{person}{R.~Kazman},
  \bibinfo{person}{J.~Grundy}, {and} \bibinfo{person}{B.~Schmerl}} (Eds.).
  \bibinfo{publisher}{Morgan Kaufmann}, \bibinfo{pages}{45--77}.
\newblock
\showISBNx{978-0-12-802855-1}
\urldef\tempurl%
\url{https://doi.org/10.1016/B978-0-12-802855-1.00003-4}
\showDOI{\tempurl}


\bibitem[\protect\citeauthoryear{Marks, Muller, Vogeli, Jung, Jazdi, and
  Weyrich}{Marks et~al\mbox{.}}{2018}]%
        {Marks18}
\bibfield{author}{\bibinfo{person}{P. Marks}, \bibinfo{person}{T. Muller},
  \bibinfo{person}{D. Vogeli}, \bibinfo{person}{T. Jung}, \bibinfo{person}{N.
  Jazdi}, {and} \bibinfo{person}{M. Weyrich}.} \bibinfo{year}{2018}\natexlab{}.
\newblock \showarticletitle{Agent Design Patterns for Assistance Systems in
  Various Domains - a Survey}. In \bibinfo{booktitle}{\emph{IEEE International
  Conference on Automation Science and Engineering (CASE)}}.
  \bibinfo{pages}{168--173}.
\newblock
\urldef\tempurl%
\url{https://doi.org/10.1109/COASE.2018.8560391}
\showDOI{\tempurl}


\bibitem[\protect\citeauthoryear{Meyer}{Meyer}{2014}]%
        {6802994}
\bibfield{author}{\bibinfo{person}{M. Meyer}.} \bibinfo{year}{2014}\natexlab{}.
\newblock \showarticletitle{Continuous Integration and Its Tools}.
\newblock \bibinfo{journal}{\emph{IEEE Software}} \bibinfo{volume}{31},
  \bibinfo{number}{03} (\bibinfo{date}{may} \bibinfo{year}{2014}),
  \bibinfo{pages}{14--16}.
\newblock
\showISSN{1937-4194}
\urldef\tempurl%
\url{https://doi.org/10.1109/MS.2014.58}
\showDOI{\tempurl}


\bibitem[\protect\citeauthoryear{Mishra and Otaiwi}{Mishra and Otaiwi}{2020}]%
        {MISHRA2020100308}
\bibfield{author}{\bibinfo{person}{A. Mishra} {and} \bibinfo{person}{Z.
  Otaiwi}.} \bibinfo{year}{2020}\natexlab{}.
\newblock \showarticletitle{DevOps and software quality: A systematic mapping}.
\newblock \bibinfo{journal}{\emph{Computer Science Review}}
  \bibinfo{volume}{38} (\bibinfo{year}{2020}), \bibinfo{pages}{100308}.
\newblock
\showISSN{1574-0137}
\urldef\tempurl%
\url{https://doi.org/10.1016/j.cosrev.2020.100308}
\showDOI{\tempurl}


\bibitem[\protect\citeauthoryear{Moreno, C\'{a}mara, Garlan, and
  Schmerl}{Moreno et~al\mbox{.}}{2015}]%
        {2786805.2786853}
\bibfield{author}{\bibinfo{person}{G. Moreno}, \bibinfo{person}{J. C\'{a}mara},
  \bibinfo{person}{D. Garlan}, {and} \bibinfo{person}{B. Schmerl}.}
  \bibinfo{year}{2015}\natexlab{}.
\newblock \showarticletitle{Proactive Self-Adaptation under Uncertainty: A
  Probabilistic Model Checking Approach}. In \bibinfo{booktitle}{\emph{10th
  Joint Meeting on Foundations of Software Engineering}}.
  \bibinfo{publisher}{ACM}, \bibinfo{pages}{1–12}.
\newblock
\showISBNx{9781450336758}
\urldef\tempurl%
\url{https://doi.org/10.1145/2786805.2786853}
\showDOI{\tempurl}


\bibitem[\protect\citeauthoryear{Muccini and Vaidhyanathan}{Muccini and
  Vaidhyanathan}{2021}]%
        {muccini2021software}
\bibfield{author}{\bibinfo{person}{H. Muccini} {and} \bibinfo{person}{K.
  Vaidhyanathan}.} \bibinfo{year}{2021}\natexlab{}.
\newblock \bibinfo{title}{Software Architecture for ML-based Systems: What
  Exists and What Lies Ahead}.
\newblock
\newblock
\showeprint[arxiv]{2103.07950}~[cs.SE]


\bibitem[\protect\citeauthoryear{Musil, Musil, Weyns, Bures, Muccini, and
  Sharaf}{Musil et~al\mbox{.}}{2017}]%
        {Musil2017}
\bibfield{author}{\bibinfo{person}{A. Musil}, \bibinfo{person}{J. Musil},
  \bibinfo{person}{D. Weyns}, \bibinfo{person}{T. Bures}, \bibinfo{person}{H.
  Muccini}, {and} \bibinfo{person}{M. Sharaf}.}
  \bibinfo{year}{2017}\natexlab{}.
\newblock \bibinfo{booktitle}{\emph{Patterns for Self-Adaptation in
  Cyber-Physical Systems}}.
\newblock \bibinfo{publisher}{Springer}, \bibinfo{pages}{331--368}.
\newblock
\showISBNx{978-3-319-56345-9}
\urldef\tempurl%
\url{https://doi.org/10.1007/978-3-319-56345-9_13}
\showDOI{\tempurl}


\bibitem[\protect\citeauthoryear{Musil, Musil, Weyns, and Biffl}{Musil
  et~al\mbox{.}}{2015}]%
        {7158500}
\bibfield{author}{\bibinfo{person}{J. Musil}, \bibinfo{person}{A. Musil},
  \bibinfo{person}{D. Weyns}, {and} \bibinfo{person}{S. Biffl}.}
  \bibinfo{year}{2015}\natexlab{}.
\newblock \showarticletitle{An Architecture Framework for Collective
  Intelligence Systems}. In \bibinfo{booktitle}{\emph{12th Working IEEE/IFIP
  Conference on Software Architecture}}. \bibinfo{pages}{21--30}.
\newblock
\urldef\tempurl%
\url{https://doi.org/10.1109/WICSA.2015.30}
\showDOI{\tempurl}


\bibitem[\protect\citeauthoryear{Musić and Hirche}{Musić and Hirche}{2016}]%
        {MUSIC201642}
\bibfield{author}{\bibinfo{person}{S. Musić} {and} \bibinfo{person}{S.
  Hirche}.} \bibinfo{year}{2016}\natexlab{}.
\newblock \showarticletitle{Classification of human-robot team interaction
  paradigms}.
\newblock \bibinfo{journal}{\emph{IFAC-PapersOnLine}} \bibinfo{volume}{49},
  \bibinfo{number}{32} (\bibinfo{year}{2016}), \bibinfo{pages}{42--47}.
\newblock
\showISSN{2405-8963}
\urldef\tempurl%
\url{https://doi.org/10.1016/j.ifacol.2016.12.187}
\showDOI{\tempurl}
\newblock
\shownote{Cyber-Physical \& Human-Systems.}


\bibitem[\protect\citeauthoryear{Naumann, Dick, Kern, and Johann}{Naumann
  et~al\mbox{.}}{2011}]%
        {GREENSOFT}
\bibfield{author}{\bibinfo{person}{S. Naumann}, \bibinfo{person}{M. Dick},
  \bibinfo{person}{E. Kern}, {and} \bibinfo{person}{T. Johann}.}
  \bibinfo{year}{2011}\natexlab{}.
\newblock \showarticletitle{The {GREENSOFT Model}: A reference model for green
  and sustainable software and its engineering}.
\newblock \bibinfo{journal}{\emph{Sustainable Computing: Informatics and
  Systems}} \bibinfo{volume}{1}, \bibinfo{number}{4} (\bibinfo{year}{2011}),
  \bibinfo{pages}{294--304}.
\newblock
\showISSN{2210-5379}
\urldef\tempurl%
\url{https://doi.org/10.1016/j.suscom.2011.06.004}
\showDOI{\tempurl}


\bibitem[\protect\citeauthoryear{{Oreizy}, {Gorlick}, {Taylor}, {Heimhigner},
  {Johnson}, {Medvidovic}, {Quilici}, {Rosenblum}, and {Wolf}}{{Oreizy}
  et~al\mbox{.}}{1999}]%
        {Oriezy1999}
\bibfield{author}{\bibinfo{person}{P. {Oreizy}}, \bibinfo{person}{M.~M.
  {Gorlick}}, \bibinfo{person}{R.~N. {Taylor}}, \bibinfo{person}{D.
  {Heimhigner}}, \bibinfo{person}{G. {Johnson}}, \bibinfo{person}{N.
  {Medvidovic}}, \bibinfo{person}{A. {Quilici}}, \bibinfo{person}{D.~S.
  {Rosenblum}}, {and} \bibinfo{person}{A.~L. {Wolf}}.}
  \bibinfo{year}{1999}\natexlab{}.
\newblock \showarticletitle{An architecture-based approach to self-adaptive
  software}.
\newblock \bibinfo{journal}{\emph{IEEE Intelligent Systems and their
  Applications}} \bibinfo{volume}{14}, \bibinfo{number}{3}
  (\bibinfo{year}{1999}), \bibinfo{pages}{54--62}.
\newblock


\bibitem[\protect\citeauthoryear{Palar, Yang, Shimoyama, Emmerich, and
  B\"{a}ck}{Palar et~al\mbox{.}}{2018}]%
        {10.1145/3205455.3205497}
\bibfield{author}{\bibinfo{person}{P.S. Palar}, \bibinfo{person}{K. Yang},
  \bibinfo{person}{K. Shimoyama}, \bibinfo{person}{M. Emmerich}, {and}
  \bibinfo{person}{T. B\"{a}ck}.} \bibinfo{year}{2018}\natexlab{}.
\newblock \showarticletitle{Multi-Objective Aerodynamic Design with User
  Preference Using Truncated Expected Hypervolume Improvement}. In
  \bibinfo{booktitle}{\emph{Genetic and Evolutionary Computation Conference}}
  (Kyoto, Japan). \bibinfo{publisher}{Association for Computing Machinery},
  \bibinfo{address}{New York, NY, USA}, \bibinfo{pages}{1333–1340}.
\newblock
\showISBNx{9781450356183}
\urldef\tempurl%
\url{https://doi.org/10.1145/3205455.3205497}
\showDOI{\tempurl}


\bibitem[\protect\citeauthoryear{Parisi, Kemker, Part, Kanan, and
  Wermter}{Parisi et~al\mbox{.}}{2019}]%
        {PARISI201954}
\bibfield{author}{\bibinfo{person}{G.~I. Parisi}, \bibinfo{person}{R. Kemker},
  \bibinfo{person}{J.~L. Part}, \bibinfo{person}{C. Kanan}, {and}
  \bibinfo{person}{S. Wermter}.} \bibinfo{year}{2019}\natexlab{}.
\newblock \showarticletitle{Continual lifelong learning with neural networks: A
  review}.
\newblock \bibinfo{journal}{\emph{Neural Networks}}  \bibinfo{volume}{113}
  (\bibinfo{year}{2019}), \bibinfo{pages}{54--71}.
\newblock
\showISSN{0893-6080}
\urldef\tempurl%
\url{https://doi.org/10.1016/j.neunet.2019.01.012}
\showDOI{\tempurl}


\bibitem[\protect\citeauthoryear{Paulovich, Oliveira, and Oliveira}{Paulovich
  et~al\mbox{.}}{2018}]%
        {Paulovich2018}
\bibfield{author}{\bibinfo{person}{F. Paulovich}, \bibinfo{person}{M.~De
  Oliveira}, {and} \bibinfo{person}{O. Oliveira}.}
  \bibinfo{year}{2018}\natexlab{}.
\newblock \showarticletitle{A Future with Ubiquitous Sensing and Intelligent
  Systems}.
\newblock \bibinfo{journal}{\emph{ACS Sensors}} \bibinfo{volume}{3},
  \bibinfo{number}{8} (\bibinfo{year}{2018}), \bibinfo{pages}{1433--1438}.
\newblock
\urldef\tempurl%
\url{https://doi.org/10.1021/acssensors.8b00276}
\showDOI{\tempurl}


\bibitem[\protect\citeauthoryear{Reussner, Goedicke, Hasselbring, Vogel-Heuser,
  Keim, and Martin}{Reussner et~al\mbox{.}}{2019}]%
        {Reussner2019}
\bibfield{author}{\bibinfo{person}{R. Reussner}, \bibinfo{person}{M. Goedicke},
  \bibinfo{person}{W. Hasselbring}, \bibinfo{person}{B. Vogel-Heuser},
  \bibinfo{person}{J. Keim}, {and} \bibinfo{person}{L. Martin}.}
  \bibinfo{year}{2019}\natexlab{}.
\newblock \bibinfo{booktitle}{\emph{Managed Software Evolution}}.
\newblock \bibinfo{publisher}{Springer Nature}.
\newblock
\showISBNx{978-3-030-13499-0}


\bibitem[\protect\citeauthoryear{Rodríguez et~al\mbox{.}}{Rodríguez
  et~al\mbox{.}}{2017}]%
        {RODRIGUEZ2017263}
\bibfield{author}{\bibinfo{person}{P. Rodríguez} {et~al\mbox{.}}}
  \bibinfo{year}{2017}\natexlab{}.
\newblock \showarticletitle{Continuous deployment of software intensive
  products and services: A systematic mapping study}.
\newblock \bibinfo{journal}{\emph{Journal of Systems and Software}}
  \bibinfo{volume}{123} (\bibinfo{year}{2017}), \bibinfo{pages}{263--291}.
\newblock
\showISSN{0164-1212}
\urldef\tempurl%
\url{https://doi.org/10.1016/j.jss.2015.12.015}
\showDOI{\tempurl}


\bibitem[\protect\citeauthoryear{Salehie and Tahvildari}{Salehie and
  Tahvildari}{2009}]%
        {1516538}
\bibfield{author}{\bibinfo{person}{M. Salehie} {and} \bibinfo{person}{L.
  Tahvildari}.} \bibinfo{year}{2009}\natexlab{}.
\newblock \showarticletitle{Self-Adaptive Software: Landscape and Research
  Challenges}.
\newblock \bibinfo{journal}{\emph{ACM Trans. Auton. Adapt. Syst.}}
  \bibinfo{volume}{4}, \bibinfo{number}{2}, Article \bibinfo{articleno}{14}
  (\bibinfo{date}{May} \bibinfo{year}{2009}), \bibinfo{numpages}{42}~pages.
\newblock
\urldef\tempurl%
\url{https://doi.org/10.1145/1516533.1516538}
\showDOI{\tempurl}


\bibitem[\protect\citeauthoryear{Schelfthout, Coninx, Helleboogh, Holvoet,
  Steegmans, and Weyns}{Schelfthout et~al\mbox{.}}{2002}]%
        {SchelfthoutKurt2002Aip}
\bibfield{author}{\bibinfo{person}{K. Schelfthout}, \bibinfo{person}{T.
  Coninx}, \bibinfo{person}{A. Helleboogh}, \bibinfo{person}{T. Holvoet},
  \bibinfo{person}{E. Steegmans}, {and} \bibinfo{person}{D. Weyns}.}
  \bibinfo{year}{2002}\natexlab{}.
\newblock \showarticletitle{Agent implementation patterns}.
\newblock \bibinfo{journal}{\emph{Workshop on Agent-Oriented Methodologies}},
  \bibinfo{pages}{119--130}.
\newblock


\bibitem[\protect\citeauthoryear{Selic}{Selic}{2020}]%
        {Selic20}
\bibfield{author}{\bibinfo{person}{B. Selic}.} \bibinfo{year}{2020}\natexlab{}.
\newblock \showarticletitle{Controlling the Controllers: What Software People
  Can Learn From Control Theory}.
\newblock \bibinfo{journal}{\emph{{IEEE} Softw.}} \bibinfo{volume}{37},
  \bibinfo{number}{6} (\bibinfo{year}{2020}), \bibinfo{pages}{99--103}.
\newblock
\urldef\tempurl%
\url{https://doi.org/10.1109/MS.2020.3006970}
\showDOI{\tempurl}


\bibitem[\protect\citeauthoryear{{Sztipanovits}, {Koutsoukos}, {Karsai},
  {Kottenstette}, {Antsaklis}, {Gupta}, {Goodwine}, {Baras}, and
  {Wang}}{{Sztipanovits} et~al\mbox{.}}{2012}]%
        {6008519}
\bibfield{author}{\bibinfo{person}{J. {Sztipanovits}}, \bibinfo{person}{X.
  {Koutsoukos}}, \bibinfo{person}{G. {Karsai}}, \bibinfo{person}{N.
  {Kottenstette}}, \bibinfo{person}{P. {Antsaklis}}, \bibinfo{person}{V.
  {Gupta}}, \bibinfo{person}{B. {Goodwine}}, \bibinfo{person}{J. {Baras}},
  {and} \bibinfo{person}{S. {Wang}}.} \bibinfo{year}{2012}\natexlab{}.
\newblock \showarticletitle{Toward a Science of Cyber–Physical System
  Integration}.
\newblock \bibinfo{journal}{\emph{Proc. IEEE}} \bibinfo{volume}{100},
  \bibinfo{number}{1} (\bibinfo{year}{2012}), \bibinfo{pages}{29--44}.
\newblock
\urldef\tempurl%
\url{https://doi.org/10.1109/JPROC.2011.2161529}
\showDOI{\tempurl}


\bibitem[\protect\citeauthoryear{Tamai}{Tamai}{2019}]%
        {Tamai2019}
\bibfield{author}{\bibinfo{person}{T. Tamai}.} \bibinfo{year}{2019}\natexlab{}.
\newblock \bibinfo{booktitle}{\emph{Key Software Engineering Paradigms and
  Modeling Methods}}.
\newblock \bibinfo{publisher}{Springer International Publishing},
  \bibinfo{address}{Cham}, \bibinfo{pages}{349--374}.
\newblock
\showISBNx{978-3-030-00262-6}
\urldef\tempurl%
\url{https://doi.org/10.1007/978-3-030-00262-6_9}
\showDOI{\tempurl}


\bibitem[\protect\citeauthoryear{Tao, Zhang, Liu, and Nee}{Tao
  et~al\mbox{.}}{2019}]%
        {8477101}
\bibfield{author}{\bibinfo{person}{F. Tao}, \bibinfo{person}{H. Zhang},
  \bibinfo{person}{A. Liu}, {and} \bibinfo{person}{A. Nee}.}
  \bibinfo{year}{2019}\natexlab{}.
\newblock \showarticletitle{Digital Twin in Industry: State-of-the-Art}.
\newblock \bibinfo{journal}{\emph{IEEE Transactions on Industrial Informatics}}
  \bibinfo{volume}{15}, \bibinfo{number}{4} (\bibinfo{year}{2019}),
  \bibinfo{pages}{2405--2415}.
\newblock
\urldef\tempurl%
\url{https://doi.org/10.1109/TII.2018.2873186}
\showDOI{\tempurl}


\bibitem[\protect\citeauthoryear{{Tavcar} and {Horv\'ath}}{{Tavcar} and
  {Horv\'ath}}{2019}]%
        {8329014}
\bibfield{author}{\bibinfo{person}{J. {Tavcar}} {and} \bibinfo{person}{I.
  {Horv\'ath}}.} \bibinfo{year}{2019}\natexlab{}.
\newblock \showarticletitle{A Review of the Principles of Designing Smart
  Cyber-Physical Systems for Run-Time Adaptation: Learned Lessons and Open
  Issues}.
\newblock \bibinfo{journal}{\emph{IEEE Transactions on Systems, Man, and
  Cybernetics: Systems}} \bibinfo{volume}{49}, \bibinfo{number}{1}
  (\bibinfo{year}{2019}), \bibinfo{pages}{145--158}.
\newblock
\urldef\tempurl%
\url{https://doi.org/10.1109/TSMC.2018.2814539}
\showDOI{\tempurl}


\bibitem[\protect\citeauthoryear{Thill, Konen, Wang, and B\"ack}{Thill
  et~al\mbox{.}}{2021}]%
        {THILL2021107751}
\bibfield{author}{\bibinfo{person}{M. Thill}, \bibinfo{person}{W. Konen},
  \bibinfo{person}{H. Wang}, {and} \bibinfo{person}{T. B\"ack}.}
  \bibinfo{year}{2021}\natexlab{}.
\newblock \showarticletitle{Temporal convolutional autoencoder for unsupervised
  anomaly detection in time series}.
\newblock \bibinfo{journal}{\emph{Applied Soft Computing}}
  \bibinfo{volume}{112} (\bibinfo{year}{2021}), \bibinfo{pages}{107751}.
\newblock
\showISSN{1568-4946}
\urldef\tempurl%
\url{https://doi.org/10.1016/j.asoc.2021.107751}
\showDOI{\tempurl}


\bibitem[\protect\citeauthoryear{Thrun and Mitchell}{Thrun and
  Mitchell}{1995}]%
        {978-3-642-79629-6-7}
\bibfield{author}{\bibinfo{person}{S. Thrun} {and} \bibinfo{person}{T.~M.
  Mitchell}.} \bibinfo{year}{1995}\natexlab{}.
\newblock \showarticletitle{Lifelong Robot Learning}. In
  \bibinfo{booktitle}{\emph{The Biology and Technology of Intelligent
  Autonomous Agents}}. \bibinfo{publisher}{Springer},
  \bibinfo{pages}{165--196}.
\newblock
\showISBNx{978-3-642-79629-6}


\bibitem[\protect\citeauthoryear{Tzafestas}{Tzafestas}{2012}]%
        {Tzafestas2012}
\bibfield{author}{\bibinfo{person}{S.G. Tzafestas}.}
  \bibinfo{year}{2012}\natexlab{}.
\newblock \bibinfo{booktitle}{\emph{Advances in intelligent autonomous
  systems}}.
\newblock \bibinfo{publisher}{Springer}.
\newblock
\showISBNx{978-94-010-6012-7}


\bibitem[\protect\citeauthoryear{Weyns}{Weyns}{2019}]%
        {DBLP:books/sp/19/Weyns19}
\bibfield{author}{\bibinfo{person}{D. Weyns}.} \bibinfo{year}{2019}\natexlab{}.
\newblock \showarticletitle{Software Engineering of Self-adaptive Systems}.
\newblock In \bibinfo{booktitle}{\emph{Handbook of Software Engineering.}},
  \bibfield{editor}{\bibinfo{person}{Sungdeok Cha}, \bibinfo{person}{Richard~N.
  Taylor}, {and} \bibinfo{person}{Kyo~C. Kang}} (Eds.).
  \bibinfo{pages}{399--443}.
\newblock
\urldef\tempurl%
\url{https://doi.org/10.1007/978-3-030-00262-6}
\showDOI{\tempurl}


\bibitem[\protect\citeauthoryear{Weyns}{Weyns}{2021}]%
        {weyns2020book}
\bibfield{author}{\bibinfo{person}{D. Weyns}.} \bibinfo{year}{2021}\natexlab{}.
\newblock \showarticletitle{Introduction to Self-Adaptive Systems: A
  Contemporary Software Engineering Perspective}.
\newblock \bibinfo{publisher}{Wiley}.
\newblock
\newblock
\shownote{ISBN 978-1-119-57494-1.}


\bibitem[\protect\citeauthoryear{Weyns, Andersson, Caporuscio, Flammini,
  Kerren, and L\"owe}{Weyns et~al\mbox{.}}{2022}]%
        {Weyns2022}
\bibfield{author}{\bibinfo{person}{D. Weyns}, \bibinfo{person}{J. Andersson},
  \bibinfo{person}{M. Caporuscio}, \bibinfo{person}{F. Flammini},
  \bibinfo{person}{A. Kerren}, {and} \bibinfo{person}{W. L\"owe}.}
  \bibinfo{year}{2022}\natexlab{}.
\newblock \showarticletitle{A Research Agenda for Smarter Cyber-Physical
  Systems}.
\newblock \bibinfo{journal}{\emph{Journal of Integrated Design and Process
  Science}} (\bibinfo{year}{2022}).
\newblock
\showISSN{1092-0617}
\urldef\tempurl%
\url{https://doi.org/10.3233/JID-210010}
\showDOI{\tempurl}


\bibitem[\protect\citeauthoryear{Weyns, B\"ack, Vidal, Yao, and
  Belbachir}{Weyns et~al\mbox{.}}{2021a}]%
        {lifelong}
\bibfield{author}{\bibinfo{person}{D. Weyns}, \bibinfo{person}{T. B\"ack},
  \bibinfo{person}{R. Vidal}, \bibinfo{person}{X. Yao}, {and}
  \bibinfo{person}{A.N. Belbachir}.} \bibinfo{year}{2021}\natexlab{a}.
\newblock \showarticletitle{Lifelong Computing}.
\newblock \bibinfo{journal}{\emph{arXiv}}  \bibinfo{volume}{abs/2108.08802}
  (\bibinfo{year}{2021}).
\newblock


\bibitem[\protect\citeauthoryear{Weyns, Bencomo, Calinescu, Camara, Ghezzi,
  Grassi, Grunske, Inverardi, Jezequel, Malek, Mirandola, Mori, and
  Tamburrelli}{Weyns et~al\mbox{.}}{2017}]%
        {10.1007/978-3-319-74183-3_2}
\bibfield{author}{\bibinfo{person}{D. Weyns}, \bibinfo{person}{N. Bencomo},
  \bibinfo{person}{R. Calinescu}, \bibinfo{person}{J. Camara},
  \bibinfo{person}{C. Ghezzi}, \bibinfo{person}{V. Grassi}, \bibinfo{person}{L.
  Grunske}, \bibinfo{person}{P. Inverardi}, \bibinfo{person}{J-M. Jezequel},
  \bibinfo{person}{S. Malek}, \bibinfo{person}{R. Mirandola},
  \bibinfo{person}{M. Mori}, {and} \bibinfo{person}{G. Tamburrelli}.}
  \bibinfo{year}{2017}\natexlab{}.
\newblock \showarticletitle{Perpetual Assurances for Self-Adaptive Systems}. In
  \bibinfo{booktitle}{\emph{Software Engineering for Self-Adaptive Systems III.
  Assurances}}, \bibfield{editor}{\bibinfo{person}{R.~de~Lemos},
  \bibinfo{person}{D.~Garlan}, \bibinfo{person}{C.~Ghezzi}, {and}
  \bibinfo{person}{H.~Giese}} (Eds.). \bibinfo{publisher}{Springer
  International Publishing}, \bibinfo{address}{Cham}, \bibinfo{pages}{31--63}.
\newblock
\showISBNx{978-3-319-74183-3}


\bibitem[\protect\citeauthoryear{Weyns, Bures, Calinescu, Craggs, Fitzgerald,
  Garlan, Nuseibeh, Pasquale, Rashid, Ruchkin, and Schmerl}{Weyns
  et~al\mbox{.}}{2021b}]%
        {WeynsBCCFGNPRRS21}
\bibfield{author}{\bibinfo{person}{D. Weyns}, \bibinfo{person}{T. Bures},
  \bibinfo{person}{R. Calinescu}, \bibinfo{person}{B. Craggs},
  \bibinfo{person}{J. Fitzgerald}, \bibinfo{person}{D. Garlan},
  \bibinfo{person}{B. Nuseibeh}, \bibinfo{person}{L. Pasquale},
  \bibinfo{person}{A. Rashid}, \bibinfo{person}{I. Ruchkin}, {and}
  \bibinfo{person}{B. Schmerl}.} \bibinfo{year}{2021}\natexlab{b}.
\newblock \showarticletitle{Six Software Engineering Principles for Smarter
  Cyber-Physical Systems}. In \bibinfo{booktitle}{\emph{{IEEE} International
  Conference on Autonomic Computing and Self-Organizing Systems, {ACSOS} 2021,
  Companion Volume, Washington, DC, USA, September 27 - Oct. 1, 2021}}.
  \bibinfo{publisher}{{IEEE}}, \bibinfo{pages}{198--203}.
\newblock
\urldef\tempurl%
\url{https://doi.org/10.1109/ACSOS-C52956.2021.00058}
\showDOI{\tempurl}


\bibitem[\protect\citeauthoryear{Weyns, Caporuscio, Vogel, and Kurti}{Weyns
  et~al\mbox{.}}{2015}]%
        {10.1145/2797433.2797497}
\bibfield{author}{\bibinfo{person}{D. Weyns}, \bibinfo{person}{M. Caporuscio},
  \bibinfo{person}{B. Vogel}, {and} \bibinfo{person}{A. Kurti}.}
  \bibinfo{year}{2015}\natexlab{}.
\newblock \showarticletitle{Design for Sustainability = Runtime Adaptation {U}
  Evolution}. In \bibinfo{booktitle}{\emph{1st International Workshop on
  Sustainable Architecture: Global collaboration, Requirements, Analysis}}
  (Dubrovnik, Cavtat, Croatia).
\newblock
\urldef\tempurl%
\url{https://doi.org/10.1145/2797433.2797497}
\showDOI{\tempurl}


\bibitem[\protect\citeauthoryear{Weyns and Iftikhar}{Weyns and
  Iftikhar}{2022}]%
        {weyns2019activforms}
\bibfield{author}{\bibinfo{person}{D. Weyns} {and} \bibinfo{person}{M.~U.
  Iftikhar}.} \bibinfo{year}{2022}\natexlab{}.
\newblock \showarticletitle{{ActivFORMS: A Formally-Founded Model-Based
  Approach to Engineer Self-Adaptive Systems}}.
\newblock \bibinfo{journal}{\emph{ACM Transactions on Software Engineering and
  Methodology}} \bibinfo{volume}{31}, \bibinfo{number}{3}
  (\bibinfo{year}{2022}).
\newblock


\bibitem[\protect\citeauthoryear{Weyns, Iftikhar, Hughes, and Matthys}{Weyns
  et~al\mbox{.}}{2018}]%
        {978-3-030-00761-4}
\bibfield{author}{\bibinfo{person}{D. Weyns}, \bibinfo{person}{U. Iftikhar},
  \bibinfo{person}{D. Hughes}, {and} \bibinfo{person}{N. Matthys}.}
  \bibinfo{year}{2018}\natexlab{}.
\newblock \showarticletitle{Applying Architecture-Based Adaptation to Automate
  the Management of Internet-of-Things}. In \bibinfo{booktitle}{\emph{Software
  Architecture}}. \bibinfo{publisher}{Springer}, \bibinfo{pages}{49--67}.
\newblock
\showISBNx{978-3-030-00761-4}


\bibitem[\protect\citeauthoryear{Weyns, Malek, and Andersson}{Weyns
  et~al\mbox{.}}{2010}]%
        {FORMS}
\bibfield{author}{\bibinfo{person}{D. Weyns}, \bibinfo{person}{S. Malek}, {and}
  \bibinfo{person}{J. Andersson}.} \bibinfo{year}{2010}\natexlab{}.
\newblock \showarticletitle{FORMS: A Formal Reference Model for
  Self-Adaptation}. In \bibinfo{booktitle}{\emph{Proceedings of the 7th
  International Conference on Autonomic Computing}} (Washington, DC, USA)
  \emph{(\bibinfo{series}{ICAC '10})}. \bibinfo{publisher}{Association for
  Computing Machinery}, \bibinfo{address}{New York, NY, USA},
  \bibinfo{pages}{205–214}.
\newblock
\showISBNx{9781450300742}
\urldef\tempurl%
\url{https://doi.org/10.1145/1809049.1809078}
\showDOI{\tempurl}


\bibitem[\protect\citeauthoryear{Wooldrige}{Wooldrige}{2009}]%
        {Wooldrige2009}
\bibfield{author}{\bibinfo{person}{M. Wooldrige}.}
  \bibinfo{year}{2009}\natexlab{}.
\newblock \showarticletitle{An Introduction to MultiAgent Systems}.
\newblock \bibinfo{publisher}{Wiley}.
\newblock
\newblock
\shownote{ISBN 978-0-470-51946-2.}


\bibitem[\protect\citeauthoryear{You, Robinson, and Vidal}{You
  et~al\mbox{.}}{2017}]%
        {you2017provable}
\bibfield{author}{\bibinfo{person}{C. You}, \bibinfo{person}{D. Robinson},
  {and} \bibinfo{person}{R. Vidal}.} \bibinfo{year}{2017}\natexlab{}.
\newblock \bibinfo{title}{Provable Self-Representation Based Outlier Detection
  in a Union of Subspaces}.
\newblock
\newblock
\showeprint[arxiv]{1704.03925}~[cs.CV]


\bibitem[\protect\citeauthoryear{Yu, Jin, and Olhofer}{Yu
  et~al\mbox{.}}{2020}]%
        {8638825}
\bibfield{author}{\bibinfo{person}{G. Yu}, \bibinfo{person}{Y. Jin}, {and}
  \bibinfo{person}{M. Olhofer}.} \bibinfo{year}{2020}\natexlab{}.
\newblock \showarticletitle{Benchmark Problems and Performance Indicators for
  Search of Knee Points in Multiobjective Optimization}.
\newblock \bibinfo{journal}{\emph{IEEE Transactions on Cybernetics}}
  \bibinfo{volume}{50}, \bibinfo{number}{8} (\bibinfo{year}{2020}),
  \bibinfo{pages}{3531--3544}.
\newblock
\urldef\tempurl%
\url{https://doi.org/10.1109/TCYB.2019.2894664}
\showDOI{\tempurl}


\bibitem[\protect\citeauthoryear{{Yu} and {Xue}}{{Yu} and {Xue}}{2016}]%
        {7433937}
\bibfield{author}{\bibinfo{person}{X. {Yu}} {and} \bibinfo{person}{Y. {Xue}}.}
  \bibinfo{year}{2016}\natexlab{}.
\newblock \showarticletitle{Smart Grids: A Cyber–Physical Systems
  Perspective}.
\newblock \bibinfo{journal}{\emph{Proc. IEEE}} \bibinfo{volume}{104},
  \bibinfo{number}{5} (\bibinfo{year}{2016}), \bibinfo{pages}{1058--1070}.
\newblock
\urldef\tempurl%
\url{https://doi.org/10.1109/JPROC.2015.2503119}
\showDOI{\tempurl}


\bibitem[\protect\citeauthoryear{Zeng, Yang, Lin, Ning, and Ma}{Zeng
  et~al\mbox{.}}{2020}]%
        {ZENG20201028}
\bibfield{author}{\bibinfo{person}{J. Zeng}, \bibinfo{person}{L. Yang},
  \bibinfo{person}{M. Lin}, \bibinfo{person}{H. Ning}, {and}
  \bibinfo{person}{J. Ma}.} \bibinfo{year}{2020}\natexlab{}.
\newblock \showarticletitle{A survey: Cyber-physical-social systems and their
  system-level design methodology}.
\newblock \bibinfo{journal}{\emph{Future Generation Computer Systems}}
  \bibinfo{volume}{105} (\bibinfo{year}{2020}), \bibinfo{pages}{1028--1042}.
\newblock
\showISSN{0167-739X}
\urldef\tempurl%
\url{https://doi.org/10.1016/j.future.2016.06.034}
\showDOI{\tempurl}


\end{thebibliography}

\end{document}